\documentclass[aps,a4paper,showpacs,preprintnumbers,amsmath,amssymb,
  floatfix,nofootinbib]{revtex4}
\usepackage{dcolumn}
\usepackage{bm}
\usepackage{amsmath,amssymb}
\usepackage{booktabs,subfigure}
\usepackage{graphicx,psfrag}
\usepackage{epsfig}
\usepackage{color} 
\usepackage{rotating}
\usepackage{slashed}
\usepackage{geometry}
\usepackage{rotating}
\usepackage[caption=false]{subfig}
\usepackage{pstricks}
\usepackage{color}
\allowdisplaybreaks[3]
\newcommand{\be}{\begin{equation}}
\newcommand{\ee}{\end{equation}}
\newcommand{\bea}{\begin{eqnarray}}
\newcommand{\eea}{\end{eqnarray}}
\newcommand{\cw}[1][{}]{\ensuremath{\cos^{#1} \theta_{W}}}
\newcommand{\sw}[1][{}]{\ensuremath{\sin^{#1} \theta_{W}}}

\numberwithin{equation}{section} 
\usepackage{epsfig}
\usepackage{dcolumn} 
\usepackage{bm} 

\def\gsim{\lower0.5ex\hbox{$\:\buildrel >\over\sim\:$}}
\def\lsim{\lower0.5ex\hbox{$\:\buildrel <\over\sim\:$}}

\begin{document}

\title{Invisible decays of low mass Higgs bosons in supersymmetric models}

\vskip 2cm

\author{
~P. N. Pandita,$^1$
~Monalisa Patra$^2$}

\affiliation{$^1$ P-30, North Eastern Hill University, 
                  Shillong 793 002, India \\
$^2$ Department of Theoretical Physics, Tata Institute of Fundamental Research, Mumbai 400 005, India %
             }
\thispagestyle{myheadings}


\vskip 5cm

\begin{abstract}
\noindent
The discovery of a 126 GeV Higgs like scalar at the LHC along with the non observation
of the supersymmetric particles, has in turn lead to constraining various supersymmetric models through
the Higgs data. We here consider the case of both the minimal supersymmetric 
standard model~(MSSM), as well its extension containing an additional chiral singlet  superfield, the 
next-to-minimal or nonminimal supersymmetric standard model~(NMSSM). 
A lot of work has been done in the context of the lightest scalar of these models
being identified as the 126 GeV state discovered at the LHC. We here however concentrate
on the case where we identify the 
second lightest Higgs boson as the $126$~GeV state discovered at the CERN
LHC and consider the invisible decays of the low mass Higgs bosons
in both MSSM and NMSSM.
In case of the MSSM, we consider $H \approx$ 126 GeV and $h \approx$ 98 GeV,
known as the non-decoupling regime, whereas in case of the NMSSM $h_2 \approx 126$ GeV, 
with $m_{h_1}$ and $m_{a_1}$ varying
depending on the parameter space. We find that in case of the MSSM with universal boundary conditions at
the GUT scale, it is not possible to have light neutralinos leading to
the decay channel $H\rightarrow \tilde{\chi}_1^0 \tilde{\chi}_1^0$. The 
invisible decay mode is allowed in case of certain $SO(10)$ and $E_6$ grand unified
models with large representations and nonuniversal gaugino masses at the GUT scale.
In case of the NMSSM, for the parameter space considered it is possible to have the 
invisible decay channel with universal gaugino masses at the GUT scale. We 
furthermore consider the most general case, with $M_1$ and $M_2$ as independent parameters
for both MSSM and NMSSM.
We isolate the regions in parameter space in both cases,
where the second lightest Higgs boson has a mass of 126 GeV and then concentrate
on the invisible decay of Higgs to lighter neutralinos. The other non-standard
decay mode of the Higgs is also considered in detail. The invisible Higgs branching ratio
being constrained by the LHC results, we find that in this case
with the second lightest Higgs being the 126 GeV state, more data from the LHC is
required to constrain the neutralino parameter space, compared to the 
case when the lightest Higgs boson is the 126 GeV state.
\end{abstract}

\maketitle

\section{Introduction}
\label{sec:intro}

A new era of particle physics has begun with the discovery of
the neutral scalar by the ATLAS~\cite{Aad:2012an} and CMS~\cite{Chatrchyan:2013lba} collaborations.
It is entirely likely that this state is the long sought after Higgs boson of the standard model (SM)
and is being pursued as a main window for new physics searches,
Though the recent results are already pointing 
towards a SM like Higgs, final conclusion can only be drawn through
a detailed study of the properties of the new boson.
These studies will indicate, whether the decay widths of the
particle are in accordance with the predictions of the SM or of its
extensions. Popular among the latter are the Minimal Supersymmetric 
Standard Model (MSSM) and the Next-to-Minimal Supersymmetric Model (NMSSM).
The Higgs sector in MSSM consists of five physical Higgs which includes two CP even Higgs ($h, H$),
one CP odd Higgs $A$ and a pair of charged Higgs ($H^\pm$).
In case of NMSSM, the  $\mu$ parameter of MSSM, is replaced by $\lambda<S>$, 
which is generated from a trilinear superpotential coupling 
$\lambda H_1H_2S$, when $S$ obtains a vacuum expectation value
$<S>$. Here $H_1$ and $H_2$ are the two Higgs doublet, whereas $S$
is the chiral singlet superfield. This term in turn leads to three CP-even Higgs bosons,
$h_{1,2,3}$, two CP-odd Higgs bosons, $a_{1,2},$ and a pair of charged 
Higgs bosons, $H^\pm$. 

The MSSM as well as the NMSSM predict the existence of a dark matter candidate,
which in large parts of the parameter space is a neutralino. If the neutralino is sufficiently 
light, Higgs decay to neutralinos will be kinematically allowed.
Such a light neutralino with the
required relic density is still supported by the recent
experimental results~\cite{Calibbi:2013poa, Kozaczuk:2013spa, Arbey:2013aba, Hagiwara:2013qya}. 
The presence of light neutralino therefore
has implications on the Higgs phenomenology, as it gives rise to 
the decay channel $h \rightarrow \tilde{\chi_1^0}\tilde{\chi_1^0}$,
i.e. invisible branching ratio. 
With the latest experimental results, 
fits are being performed to check how much deviation is allowed
by the recent data, in order to take into account new physics scenarios.
The invisible Higgs decay width
has been constrained by various groups by performing fits
of the signal strengths in various search channels using the
latest LHC Higgs data. The results of some of the most recent global fits:
\begin{enumerate}
\item Considering the Higgs couplings to the quarks, leptons and vector bosons to be free, \\
BR$(h \rightarrow \tilde{\chi}_1^0 \tilde{\chi}_1^0) <$ 0.16 (0.38)
at 68\% (95\%) CL~\cite{Belanger:2013xza}.
\item With the assumption that the Higgs coupling to fermions and gauge bosons are SM like, and
the only new physics is from the Higgs invisible decay width, \\
BR$(h \rightarrow \tilde{\chi}_1^0\tilde{\chi}_1^0) <$ 0.52
at 68\% CL~\cite{Djouadi:2013qya}.
\end{enumerate}
Direct search for invisible decaying Higgs produced in association
with a $Z$ boson has been carried out by the ATLAS and 
CMS collaborations in LHC. They have, in turn, placed limits
on the branching fraction of the Higgs boson to invisible
particles, with the branching fraction greater than  65\% 
and 75\% excluded at 95\% CL by ATLAS~\cite{Aad:2014iia, ATLAS:2013pma}
and CMS~\cite{CMS:1} respectively. The CMS collaboration
has also carried out a similar search for invisible branching
ratio of the Higgs boson produced in the vector boson fusion
process and have placed an upper limit of 69\%~\cite{CMS:2}.
\footnote{In our analysis we will use the most stringent limit
obtained on the invisible branching ratio to be less than 38\%~\cite{Belanger:2013xza}.}

Direct searches for SUSY particles at the LHC so far have come empty handed.  Furthermore, 
several analyses based on simple versions of the MSSM and other models have ruled out significant
regions of the parameter space. However, the parameter space under more general assumptions still 
remains largely unexplored. One  possibility of exploring these regions is to ask under what conditions 
the 126 GeV state corresponds to the neutral higgs particles in the spectrum of the model.
Popular among these is the case where the 
lightest Higgs boson of MSSM ($h$) and NMSSM ($h_1$) is identified with the state discovered
at LHC at 126 GeV and has been studied in great detail. In~\cite{Ananthanarayan:2013fga}, 
the authors studied the decay of  this lightest Higgs boson into neutralinos in 
these low energy supersymmetric models. The neutralino
sector of these models were then constrained, from the limits
on the invisible decay width. The regions of the parameter
space where the lightest Higgs boson has a mass of around 126 GeV
was isolated and then the regions where this Higgs can decay into 
light neutralinos were studied in details. It was found that it was not possible
to have a massless neutralino in MSSM both in case of universal and nonuniversal
gaugino masses at the GUT scale, except for some higher representation of $E_6$.
In case of NMSSM although it was possible to have a massless neutralino with
universal gaugino masses at the GUT scale, it was not possible to obtain $M_{h_1}$ = 126 GeV
and simultaneously have massless neutralino or $M_{\tilde{\chi}_1^0} \leq M_{h_1}/2$.
The results were therefore obtained by considering the 
soft gaugino masses $M_1$ and $M_2$ as two independent parameters,
unconstrained by grand unification. In case of NMSSM, for certain
regions of the parameter space there were additional nonstandard decay
channels like $h_1 \rightarrow a_1a_1,~Za_1$. The composition of the $\tilde{\chi}_1^0$
was important in determining the invisible branching ratio, with some regions of the 
parameter space allowing large invisible branching ratio. 

Note that there are no a priori
reasons to believe that the 126 GeV boson is the lightest Higgs
boson. In the MSSM either $h$ and $H$ can be identified with the discovered 126 GeV boson,
with either $h \approx$ 126 GeV, $H \geq$ 800 GeV or $H \approx$ 126 GeV,
$M_h < M_H$. Analogously for the  NMSSM there can be many choices,
($a$) $h_1 \approx$ 126 GeV, and $M_{h_{2,3}} > M_{h_1}$ 
($b$)  $h_2 \approx$ 126 GeV, and $M_{h_{3}} > M_{h_2},~M_{h_{1}} < M_{h_2}$.
The mass of the CP-odd Higgs Bosons varies in the range of 4 GeV to TeV (to
be discussed later), whereas 
the charged Higgs boson are very massive with masses of the order 1 TeV. 
This scenario with $M_H \approx$ 126 GeV is mainly in light of the observed 
LEP excess~\cite{Barate:2003sz} in the
$e^+e^- \rightarrow Zh \rightarrow Zb\bar{b}$ channel around
$M_{b\bar{b}} \approx$ 100 GeV, which indicates that there may be a lighter
Higgs boson less than 100 GeV. 

Here we take this possibility seriously and ask under what circumstances this is 
realized, and to what extent the measured properties allow this scenario to survive.
Crucial to this is the possibility that the uncertainty in the width is saturated by 
invisible decays, rendered possible when there are states lighter than 63 GeV in the spectrum.
In this work we firstly delineate regions of the parameter space, which give
rise to two light Higgs and then study in details the branching ratio of the 126 GeV
Higgs to non standard SM particles. This second lightest 126 GeV Higgs can decay to
a pair of lightest neutralinos as well as
to a pair of lightest Higgs in some regions of parameter space. We consider both
the cases here. Moreover in case of NMSSM, the decay of $h_2$ to a pair of $a_1$
is also kinematically allowed in some regions. Considering the limit on the 
invisible branching ratio from the experiments and global fits, the 
neutralino sector of these supersymmetric models are then constrained accordingly.
In most studies, the parameter space of the models are constrained
with the assumption of the universality of the gaugino masses at the GUT scale.
Note the gaugino masses need not be universal at the GUT scale.
This happens when the SM gauge group is embedded in a grand unified gauge group.
The phenomenology of the neutralinos at the weak scale, is then affected via the 
renormalization group evolution of these gaugino mass parameters.
At the weak scale, there will be a possibility of massless neutralinos~\cite{Gogoladze:2002xp}
depending on the gaugino masses at the GUT scale. We find that for the parameter space
which allows a 126 GeV, the second lightest
Higgs boson in MSSM, analogous to the case where the 126 GeV state was identified 
with the lightest Higgs boson,
it is not possible to have a massless neutralino
with universal gaugino mass parameters $M_1$ and $M_2$.
The result holds even for nonuniversal gaugino mass parameters 
except for a higher dimensional representation of $E_6$. 
Nevertheless, the decay of Higgs to lightest neutralinos ($M_{\tilde{\chi}_1^0} \leq M_{H}/2$)
is allowed for some representations in case of MSSM
with nonuniversal gaugino masses at the GUT scale.
Analogous in the case of the NMSSM, it is not possible to obtain $M_{h_2}$ = 126 GeV and 
simultaneously have massless neutralinos, with universal gaugino masses at the GUT scale
but the decay of the Higgs ($h_2$) to the lightest neutralinos is allowed 
($M_{\tilde{\chi}_1^0} \neq 0$ and $M_{\tilde{\chi}_1^0} \leq M_{h_2}/2$).
This assumption of the GUT relation between $M_1$ and $M_2$, is biased to
a particular scenario, so we do not consider the universality assumption on the 
gaugino mass parameters and rather treat $M_1$, $M_2$ as two independent 
parameters. 

In view of the considerations above, we have now considered the possibility of the 
second lightest neutral Higgs to be the 126 GeV state discovered by the LHC, in some versions of the
MSSM and the NMSSM along with its invisible decays due to the presence of light neutralinos
or other light states present in the spectrum. Thus, the outline of the draft is
as follows. In Sec.~\ref{sec:mssm}
we study the existence of a massless neutralino or a neutralino
with  mass less than the half the mass of 126 GeV Higgs boson in the context
of MSSM, with appropriate boundary conditions as dictated by grand unification
based on $SU(5)$, $SO(10)$ and $E_6$ gauge groups. The relevant experimental
constraints on the lightest Higgs ($m_h < m_H$, with $m_H \approx$ 126 GeV)
as well as other SUSY particles is considered. The decay of the lightest
as well as the second lightest Higgs to neutralinos is considered in 
SubSec.~\ref{subsec:mssm}. The Higgs sector in case of NMSSM is investigated
in detail in Sec.~\ref{sec:nmssm}. We first isolate the parameter space
which supports $h_2$ in the appropriate mass window 
123-127 GeV. Then the invisible decay of the second lightest Higgs boson
along with the other nonstandard decay modes is considered in detail.
The dominant decay mode of the lightest Higgs boson $h_1$ is also considered 
in this section. Finally, we summarize our results in Sec.~\ref{sec:conclusion}.

\section{Higgs and Neutralino Sector in MSSM}\label{sec:mssm}
Let us begin by recalling that the Higgs sector in MSSM has five physical mass eigenstates,
two CP even and one CP odd neutral along with
a pair of charged scalar bosons. The Higgs spectrum at 
tree level is completely
determined by two independent parameters $M_A$ and $\tan \beta$,
where $\tan \beta$
is the ratio of the vacuum expectation value of the Higgs field
and $M_A$ is the mass of the pseudoscalar Higgs. 
In addition to this, the MSSM Higgs sector also depends
on the stop masses along with the stop mixing parameter $X_t$,
when the radiative corrections are taken into account. Here
$X_t = A_t - \mu/\tan \beta$, with $A_t$ as the trilinear Higgs-stop
coupling and $\mu$ is the Higgsino mass parameter. 
The Higgs sector of the MSSM, with a Higgs Boson of mass $\approx$ 126 GeV, 
and with SM like cross sections and branching fractions, 
can be broadly divided into two distinct regimes
depending on the magnitude of $M_A$.
\begin{itemize}
 \item The decoupling regime, where $M_A~\gg~M_Z$.  In this case the lightest CP even
boson $h$ has mass around 126 GeV, whereas all the 
others $H$, $A$ and $H^\pm$ are almost degenerate and have
mass equal to $M_A$.  
\item The non-decoupling regime is the one where $M_A \leq 130$ GeV.
The heavy CP even state $H$ is SM like, whereas the other neutral
bosons are almost degenerate in mass $M_h \equiv M_A$. The mass
of $h$ and $A$ can vary from the $Z$ boson mass to the heavy
neutral state $H$, depending on the value of $M_A$ and $\tan \beta$.
The charged state $H^\pm$ will be slightly heavier, but still
light enough to be detected in the Large Hadron Collider. Moreover in this
regime, with the mass of $H$ being around 126 GeV, the mass of the other
gauge bosons like $h,~A$ and $H^\pm$ are bounded from above.
\end{itemize}
The LEP collaborations have placed lower bounds on the masses of the neutral
Higgs Boson $M_A$ and the lightest scalar $h$~\cite{Schael:2006cr}.
The lower bounds on $M_h$ and $M_A$ are usually obtained from the upper bound on 
the cross section $\sigma (e^+e^-\rightarrow Zh)$ 
and $\sigma(e^+e^-\rightarrow Ah)$. This in turn has led to
values of $M_h$ and $M_A$ less than 92.9 GeV and 93.4 GeV 
being excluded at 95\% C.L. Along with it values of 
$\tan \beta$ between 0.7 and 2 are also excluded.
Recent searches of the extra Higgs boson at the LHC have put new bounds on $\tan \beta$ as a function of
$M_A$~\cite{Chatrchyan:2012vp, Aad:2012yfa, CMS:gya, ATLAS:2012dsy}.  CMS
data~\cite{CMS:gya} has excluded regions of $\tan\beta$ above 6 for $M_A$
below 250 GeV in the $m_h^{\rm{max}}$ scenario.

\begin{figure}[htb]
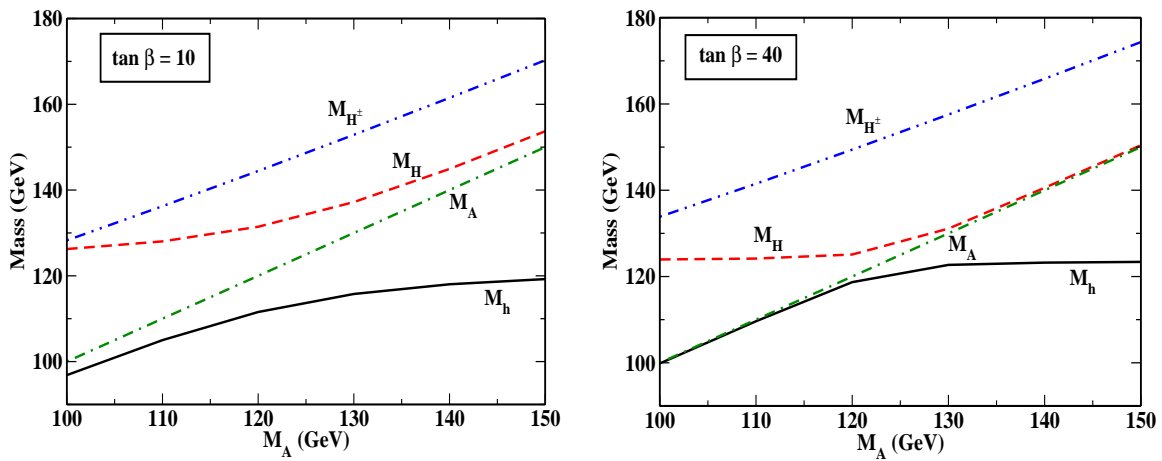

\vspace{0.65cm}
\centering 
\includegraphics[width=7.3cm, height=6cm]{tanbeta_10.eps}
\hspace{0.3cm}
\includegraphics[width=7.3cm, height=6cm]{tanbeta_40.eps}
\caption{The masses of MSSM Higgs bosons as a function of $M_A$ for two distinct values of
$\tan \beta$.}
\label{fig:ma_dependence}
\end{figure}

The main focus in this present work will be on the non-decoupling regime, and as a result, 
we would like to make some observations on the value of $M_A$ and
$\tan \beta$ chosen for our analysis. Since this regime is mainly 
characterized by the pseudoscalar mass being less than 150 GeV,
we plot in Fig.~\ref{fig:ma_dependence} the MSSM Higgs boson mass as a function of $M_A$,
for two different values of $\tan \beta$. The other SUSY parameter $A_t$, which affects 
the Higgs sector is fixed assuming  maximal stop mixing. 
\noindent 
It can be seen from the left hand side of Fig.~\ref{fig:ma_dependence}, in case of
$\tan \beta$ = 10, for $M_A \leq$ 120 GeV, $M_H$ is around 126 GeV, whereas
$M_A$ and $M_h$ are almost degenerate. The maximal value of $M_A$
which allows for $M_H$ around 126 GeV is 130 GeV. The right hand side of 
Fig.~\ref{fig:ma_dependence} shows the masses of the Higgs bosons as 
a function of $M_A$ for $\tan \beta$ = 40. We see that for our
choice of SUSY parameters, for $M_A$ around 125 GeV, the three neutral Higgs bosons
have comparable masses, $M_H \simeq M_h \simeq M_A$. 
This special case where the Higgs masses are close to each other, is called the intense coupling
scenario. 
The LHC phenomenology of this scenario has been studied in details 
in the past~\cite{Boos:2002ze, Boos:2003jt, Christensen:2012ei}. 
It has been known that in this intense coupling scenario
the neutral bosons ($h,~H$) couplings to the gauge bosons are suppressed
with respect to the SM, since $A$ does not couple to the gauge bosons.
Furthermore, the neutral bosons in this case will mainly decay to the down type fermions, 
due to the enhancement of its coupling. The recent LHC results being favoring a
SM like Higgs along with large values of $\tan\beta$ being disfavored by the 
CMS data, we have concentrated on the case with $M_A \approx$  100 GeV and 
small values of $\tan \beta$,
in order to have the cross section times branching ratio of the Higgs to any SM 
particle in agreement with the recent LHC results. The value of $\tan \beta$ 
is tuned along with the parameter $A_t$, so that  $M_H$ is in the range 
124 $<~M_H~<$ 127 GeV, and $M_h \simeq$ 97 GeV. 

In the past, various studies have shown that certain regions of the parameter
space of MSSM, allow a Higgs boson with mass 126 GeV both in the
decoupling and non-decoupling regime, satisfying the
LHC constraints. For most of the 
allowed parameter space, the Higgs decay to the lightest neutralinos
is kinematically allowed, leading to invisible decay modes. It will
therefore be very important to study the couplings of the newly 
discovered particle at high precision. As mentioned before
global fits have been performed on the couplings of the newly
discovered particle, in order to place upper bounds on the invisible
decay width. Taking into account these bounds, the parameter space of 
these new physics scenarios can be further constrained, since the regions giving a 
large invisible Higgs decay branching ratio will be in conflict with the
experiments. This was earlier done in the context of
MSSM, in the decoupling scenario, where it was found that
large regions in $\mu-M_1$ parameter space was disfavored~\cite{Dreiner:2012ex}
by the bounds on the invisible Higgs decay width, for different values of $\tan \beta$.
In this work, we have investigated further to see whether the same result 
holds in the non-decoupling scenario, assuming $H$
to be the 126 GeV boson. 

Before proceeding further we give a brief review 
of the neutralino sector in the MSSM.
The physical mass eigenstates are obtained after the electroweak symmetry breaking,
from the diagonalization of the neutralino 
mass  matrix~\cite{Bartl:1989ms,Haber:1984rc}, with
the neutralinos being an admixture of the 
fermionic partners of the two Higgs doublets, $H_1$ and $H_2$,
and the fermionic partners of the neutral gauge bosons. 
\begin{eqnarray}
\label{mssmneut}
M_{\mathrm{MSSM}} =
\begin{pmatrix}
M_1 & 0   & - M_Z \sw \cos\beta & \phantom{-}M_Z\sw \sin\beta \\
0   & M_2 & \phantom{-} M_Z \cw \cos\beta  & -M_Z \cw\sin\beta \\
 - M_Z \sw \cos\beta &\phantom{-} M_Z \cw \cos\beta  & 0 & -\mu\\
\phantom{-}M_Z\sw \sin\beta& -M_Z \cw\sin\beta & -\mu & 0
\end{pmatrix},
\end{eqnarray}
\noindent where $M_1$ and $M_2$ are the $U(1)_Y$ and 
the $SU(2)_L$ soft supersymmetry breaking gaugino
mass parameters, $\mu$ is the Higgs(ino) mass parameter, $M_Z$ is the
$Z$ boson mass, $\theta_W$ is the weak mixing angle and 
$\tan\beta=v_2/v_1$ is the ratio of the vacuum expectation values 
of the neutral components of the 
two Higgs doublet fields $H_1$ and $H_2$. 
Since we are concentrating on the Higgs invisible decay mode, 
the light neutralino eigenstate of the neutralino mass matrix~(\ref{mssmneut})
is favorable. Therefore we consider the limiting case of the massless neutralino, 
which, at the tree level, arises when the determinant of the 
mass matrix (\ref{mssmneut}) is zero.  
This in turn leads to the condition~\cite{Gogoladze:2002xp}
\begin{equation}
\mu \left[M_Z^2 \sin 2\beta \left(M_1\cos ^2 \theta_W + M_2 \sin ^2 
\theta_W \right)-M_1M_2 \right] = 0.
\label{det_mssm}
\end{equation}

\noindent 
The chargino mass lower bounds from the LEP experiments~\cite{ALEPH:2005ab},
excludes the solution $\mu$ = 0,
\begin{equation}
|\mu |, ~ ~ M_2 \geqslant 100~{\rm GeV}.
\label{mu_bound}
\end{equation}
Therefore the other possible solution to (\ref{det_mssm}) is
\begin{equation}
M_1=\frac{M_2 M_Z^2 \sin ^2 \theta_W \sin 2\beta}
{\mu M_2-M_Z^2 \cos ^2 \theta_W \sin 2\beta}.
\label{M1_mssm}
\end{equation}
\noindent 
In order to get a massless neutralino,
for fixed values of $\mu, M_2$ and $\tan \beta$, one can find a value of $M_1$ consistent with
(\ref{M1_mssm}). 

In the earlier work~\cite{Ananthanarayan:2013fga}
it was found that it is not 
possible to have a massless neutralino, both with the gaugino parameters
being universal and non universal at the GUT scale, except for 
some higher representation of $E_6$. The light neutralino with mass less than 
half the mass of the Higgs boson, is still not ruled out by the current experiments.
It is seen that for the models, with the ratio
of $M_1/M_3$ $<$ 1/28, the invisible decay of Higgs to the lightest neutralinos is allowed
and holds true for both the coupling and the non-decoupling regimes of MSSM. 
This is mainly by taking into account the constraint on the gluino mass 
($M_{\tilde{g}} \approx M_3$) $>$ 1.3 TeV from the LHC experiments, and
the other gaugino mass parameters being relatively fixed from the boundary
conditions at the electroweak scale. As can be seen from Ref.~\cite{Ananthanarayan:2013fga},
this condition on the ratio $M_1/M_3$ is satisfied by some of the higher dimensional
representation of $SO(10)$ and $E_6$.

\subsection{Decay of Higgs to Neutralinos in the MSSM}\label{subsec:mssm}

In this section we mainly concentrate on constraining the Higgs parameter
space in case of MSSM, from the Higgs invisible decay width.
One of the main assumptions that go into limiting 
the parameter space of these models is the universality of the gaugino 
mass parameters at the GUT scale. The LEP constraint on the charginos,
has led to lower bound on the lightest neutralino mass 
\begin{equation}
M_{\tilde{\chi}_1^0} > 46~{\rm GeV}
\end{equation}
at 95\% C.L. in the context of the MSSM, assuming universal gaugino
masses at the GUT scale~\cite{Abdallah:2003xe}. 
The gaugino mass parameters need not be universal at the GUT scale,
therefore the phenomenology of the neutralinos in
all these cases will be affected depending on the renormalization group evolution of
the gaugino mass parameters. We do not consider any specific representations, but instead
consider a more generic case with $M_1$ and $M_2$ as independent parameters,
in the non-decoupling scenario. With this consideration the lightest neutralino so obtained
will be bino like, because the chargino mass bounds from LEP has already set lower
limits on $M_2$ and $\mu$. 

In MSSM, the decay width of the CP even neutral scalar bosons, to a pair of
lightest neutralinos can be written as~\cite{Griest:1987qv}
\begin{eqnarray}
\Gamma (h \rightarrow \tilde{\chi}_1^0 \tilde{\chi}_1^0) 
&=& \frac{G_F M_W^2 M_h}{2 \sqrt{2} \pi}(1- 4M^2_{\tilde{\chi}^0_1}/M_h^2)^{3/2}
\left[(Z_{12} - \tan{\theta_W} Z_{11})
(Z_{13}\sin \alpha + Z_{14}\cos \alpha)\right]^2, \label{higgs_decay1} \\
\Gamma (H \rightarrow \tilde{\chi}_1^0 \tilde{\chi}_1^0) 
&=& \frac{G_F M_W^2 M_H}{2 \sqrt{2} \pi}(1- 4M^2_{\tilde{\chi}^0_1}/M_H^2)^{3/2}
\left[(Z_{12} - \tan{\theta_W} Z_{11})
(Z_{13}\cos \alpha - Z_{14}\sin \alpha)\right]^2, \label{Higgs_decay1}
\end{eqnarray}
where $Z_{ij}$ are the elements of the matrix $Z$ which diagonalizes 
the neutralino mass matrix, and $\alpha$ is the mixing angle in the $CP$ 
even Higgs sector. The above (\ref{higgs_decay1}),~(\ref{Higgs_decay1}) shows that the invisible
branching ratio requires $\tilde{\chi}_1^0$ to be a mixed state, with both
gaugino and higgsino contribution. The invisible decay of the Higgs
though favoured by a large higgsino fraction neutralino,
will be mainly constrained by the $Z$ invisible decay width. The $Z$ width
to a pair of lightest neutralinos is given by~\cite{Heinemeyer:2007bw}
\begin{equation}
 \Gamma (Z \rightarrow \tilde{\chi}_1^0 \tilde{\chi}_1^0) =
 \frac{G_F M_Z^3}{6 \sqrt{2} \pi}(1- 4M^2_{\tilde{\chi}^0_1}/M_Z^2)^{3/2}
(Z_{13}^2 - Z_{14}^2).
\end{equation}
The invisible decay width of $Z$ to a pair of lightest 
neutralinos is restricted to
\begin{equation}\label{z_width}
 \Gamma(Z^0\rightarrow \tilde{\chi}_1^0 \tilde{\chi}_1^0) < 3~{\rm MeV}.
\end{equation}
at 95\% C.L. by the LEP collaborations~\cite{ALEPH:2005ab}.

The mass bound on the lightest chargino $M_{\tilde{\chi}^\pm} >$ 94 GeV
from the LEP experiments is taken into account~\cite{Abdallah:2003xe}. 
The results are presented for a fixed value
of $M_2$ = 200 GeV, with the parameters $\mu$ and $M_1$ being varied. The other
SUSY parameters like the squarks and gluinos are fixed to masses around 1 TeV
in accordance with the latest LHC results. 
The masses of the sleptons are taken to be greater than 500 GeV.
Since we are considering the Higgs decay to the lightest neutralino pair,
which are also one of the leading dark matter candidates, it
will be necessary to check whether the 
kinematically allowed parameter region in the $\mu - M_1$ plane also gives the
correct relic density as  measured from WMAP~\cite{Bennett:2003ca, Spergel:2003cb},
i.e. 0.0925 $< \Omega h^2 <$ 0.1287.
In this work, the computation of the relic density has been performed with
micrOMEGAs 3.2~\cite{Belanger:2004yn}, along with the production and decays of the SUSY particles
being computed with CalcHEP~\cite{Belyaev:2012qa}.

Apart from the LHC constraints,
the constraints from ($g$-2) of the muon and other flavour constraints such as $b \rightarrow s \gamma$
and $B_s \rightarrow \mu^+ \mu^-$ are also taken into account, which are implemented
within CalcHEP.
\begin{figure}[htb]
\vspace{0.65cm}
\centering 
\includegraphics[width=7.3cm, height=6cm]{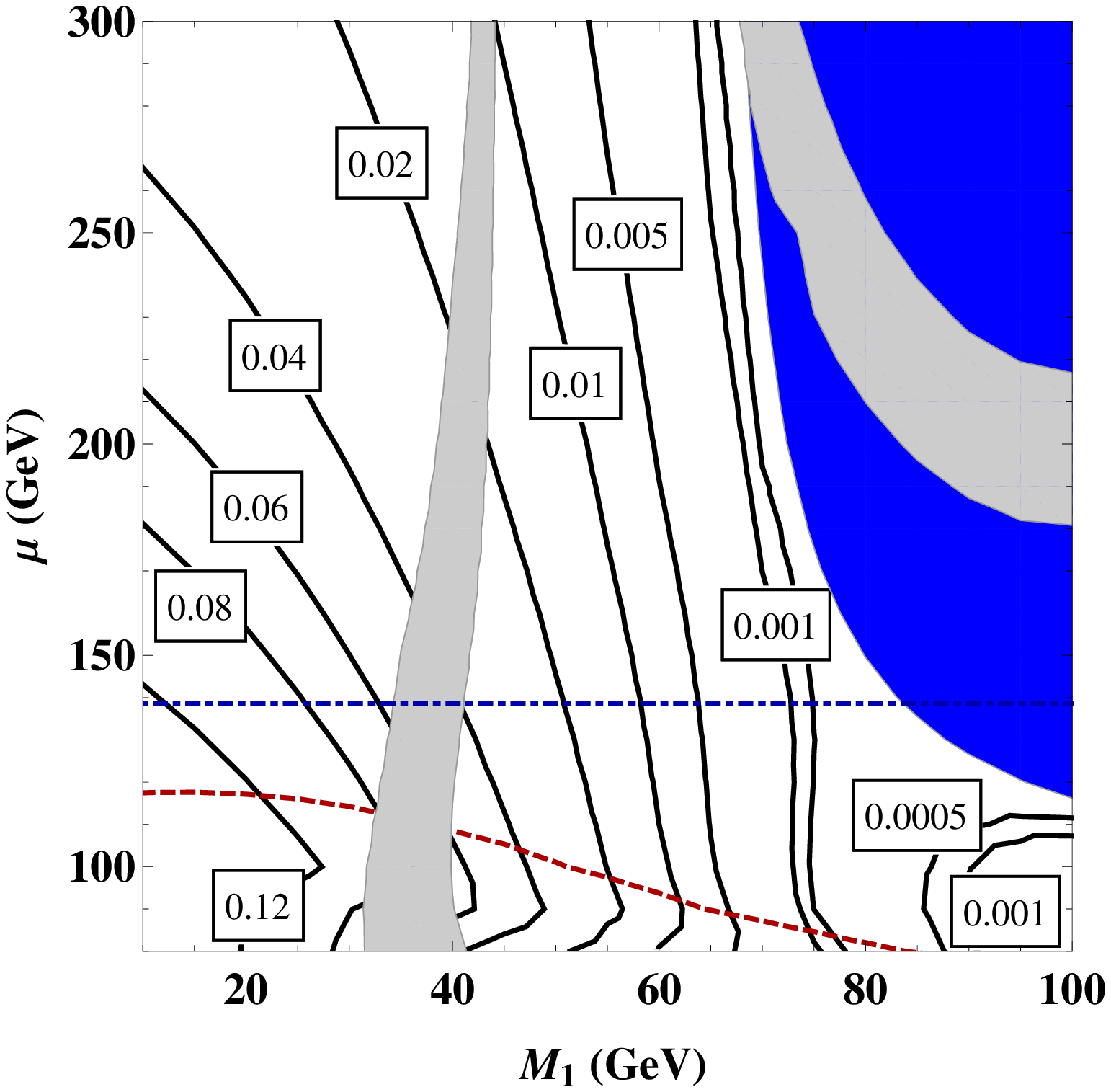}
\hspace{0.3cm}
\includegraphics[width=7.3cm, height=6cm]{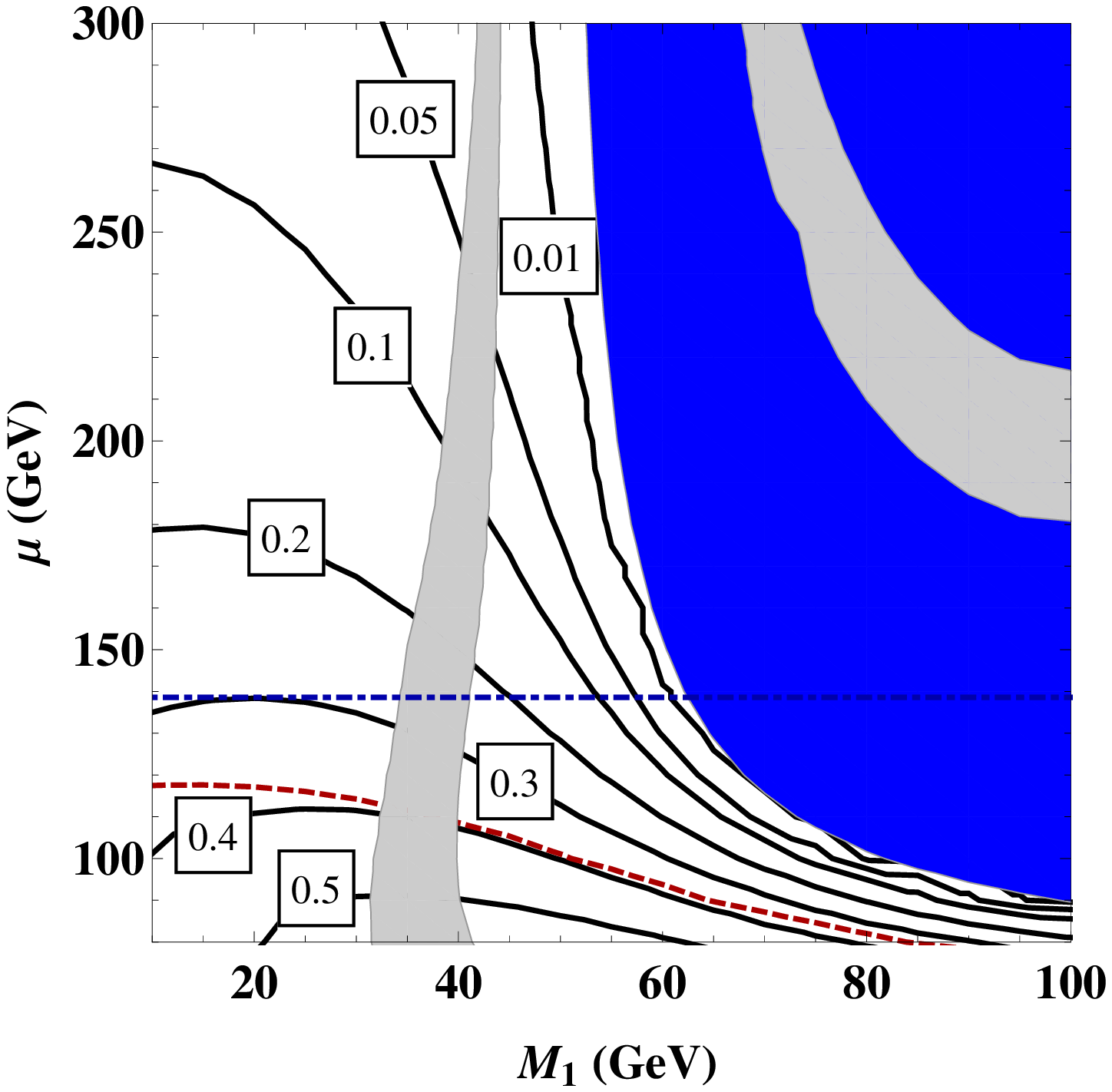}
\caption{The contours of constant branching ratio of ($H \rightarrow \tilde{\chi}_1^0 \tilde{\chi}_1^0$)
(left) and ($h \rightarrow \tilde{\chi}_1^0 \tilde{\chi}_1^0$) (right) 
for a fixed value of $\tan \beta$ = 6 and $M_2$ = 200 GeV. The region shaded in Blue is the region
kinematically not allowed, whereas the Grey shaded region is the one with the relic density
within the experimental limits.}
\label{fig:higgs_mssm}
\end{figure}

We show in Fig.~\ref{fig:higgs_mssm}, the branching ratio of both CP even Higgs
($H$ (left), $h$ (right)) to the lightest neutralinos. The regions with
large $\mu$ and $M_1$ values as expected give rise to massive neutralinos, 
with $M_{\tilde{\chi}^1_0} > M_{H,h}/2$, and is therefore kinematically not allowed 
and is shaded in Blue. The neutralino obtained by considering $M_1$ and $M_2$ as independent
parameters is mostly a bino like as the LEP mass bound on chargino has already
placed lower bound on $M_2$ and $\mu$. In the bino limit, the process which 
mainly contributes to the relic density is the one mediated by a $t$- channel sfermion.
We have calculated the relic density for different slepton masses varying from 150-1000 GeV.
There being no significant change in the allowed parameter space, we have presented
the results for sfermion mass of 500 GeV.
It is seen that for most of the allowed parameter region
of $\mu - M_1$, taking into account the LEP bound of the chargino 
mass [Blue-DotDashed] and the invisible decay width
of the $Z$ boson [Red-Dashed], the invisible branching ratio of the Higgs $H$ is 
still too small to be probed at the LHC. The area below the 
Blue-DotDashed and Red-Dashed lines, are excluded from the 
LEP bound of the chargino and the invisible $Z$ decay width
respectively.
The parameter space can not be constrained by the
latest limits on the invisible decay width from the LHC fits. This is in contrast to the situation 
$M_h \approx 126$ GeV, where most of the $\mu -M_1$ parameter space was constrained
from the bounds on the invisible branching ratio of the 
Higgs (h)~\cite{Dreiner:2012ex, Ananthanarayan:2013fga}. Moreover in the non-decoupling
scenario as can be seen from the right hand side of Fig.~\ref{fig:higgs_mssm}, the 
BR($h \rightarrow \tilde{\chi}_1^0 \tilde{\chi}_1^0$) is small compared to the 
decoupling case, due to an enhanced coupling to the $b$ quarks. The enhancement is mainly
due to the $\sin\alpha$ term, in the coupling of $h$ to a pair of $b$ quarks,
which is sensitive to the parameter $M_A$. 

The shape of the contours in the left plot of Fig.~\ref{fig:higgs_mssm},
can be understood from the fact that, the 
BR($H \rightarrow \tilde{\chi}_1^0 \tilde{\chi}_1^0$) decreases for increasing
$\mu$, due to the increase in neutralino mass. 
The dip in the contours for $\mu$ around 100 GeV is due to the fact that for
a particular value of $M_1$, after $\mu$ decreases to a
certain value, the other decay modes of Higgs such as 
$h\rightarrow \tilde{\chi}_1^0 \tilde{\chi}_2^0, \tilde{\chi}_1^0 \tilde{\chi}_3^0, 
\tilde{\chi}_1^+ \tilde{\chi}_1^-$ open up, leading to a decrease in the invisible BR.
Most of the parameter space for $\mu<$ 140 GeV, is however excluded by the 
chargino mass bound of 110 GeV. The same argument holds for $h$, the right plot of
Fig.~\ref{fig:higgs_mssm}.
We finally list in Table~\ref{tab:mssm_higgs_br}
the branchings of $h,~H$ to different final states for our parameter choice of $M_A$ = 105 GeV,
$\tan \beta$ = 6, $M_2$ = 100 GeV, $M_1$ = 50 GeV and $\mu$ = 130 GeV. The BR of decay to neutralinos
changes with the change of $\mu$ and $M_1$ as discussed before.
\begin{table}
\begin{center}
\begin{tabular}{||c|c|c||}\hline
  final states & $H$ branchings &$h$ branchings \\ \hline \hline
    & $M_H$ = 125 GeV &$M_h$ = 97 GeV \\ \hline \hline
$l,L (e,\mu, \tau)$ &0.089 &0.074 \\
$bb$ &0.841 &0.735 \\
$cc$ &0.004 &$3.5\times10^{-4}$  \\
$GG$ &0.016   &$1.1\times10^{-3}$  \\
$AA$ &$1.7\times10^{-4}$ &$3.3\times10^{-5}$ \\
$W^+W^-$ &0.024 &$6.5\times10^{-5}$ \\
$ZZ$ &$2.9\times10^{-3}$ &   \\
$\tilde{\chi}_1^0 \tilde{\chi}_1^0$ &$0.023$ &$0.191$ \\ \hline 
\end{tabular}
\caption{Branching ratios of both $h, H$ to various decay channels,
with our parameter choice of $M_A$ = 105 GeV,
$\tan \beta$ = 6, $M_2$ = 200 GeV, $M_1$ = 50 GeV and $\mu$ = 130 GeV.}
\label{tab:mssm_higgs_br}
\end{center}
\end{table}
\begin{figure}[htb]
\vspace{0.65cm}
\centering 
\includegraphics[width=7.3cm, height=6cm]{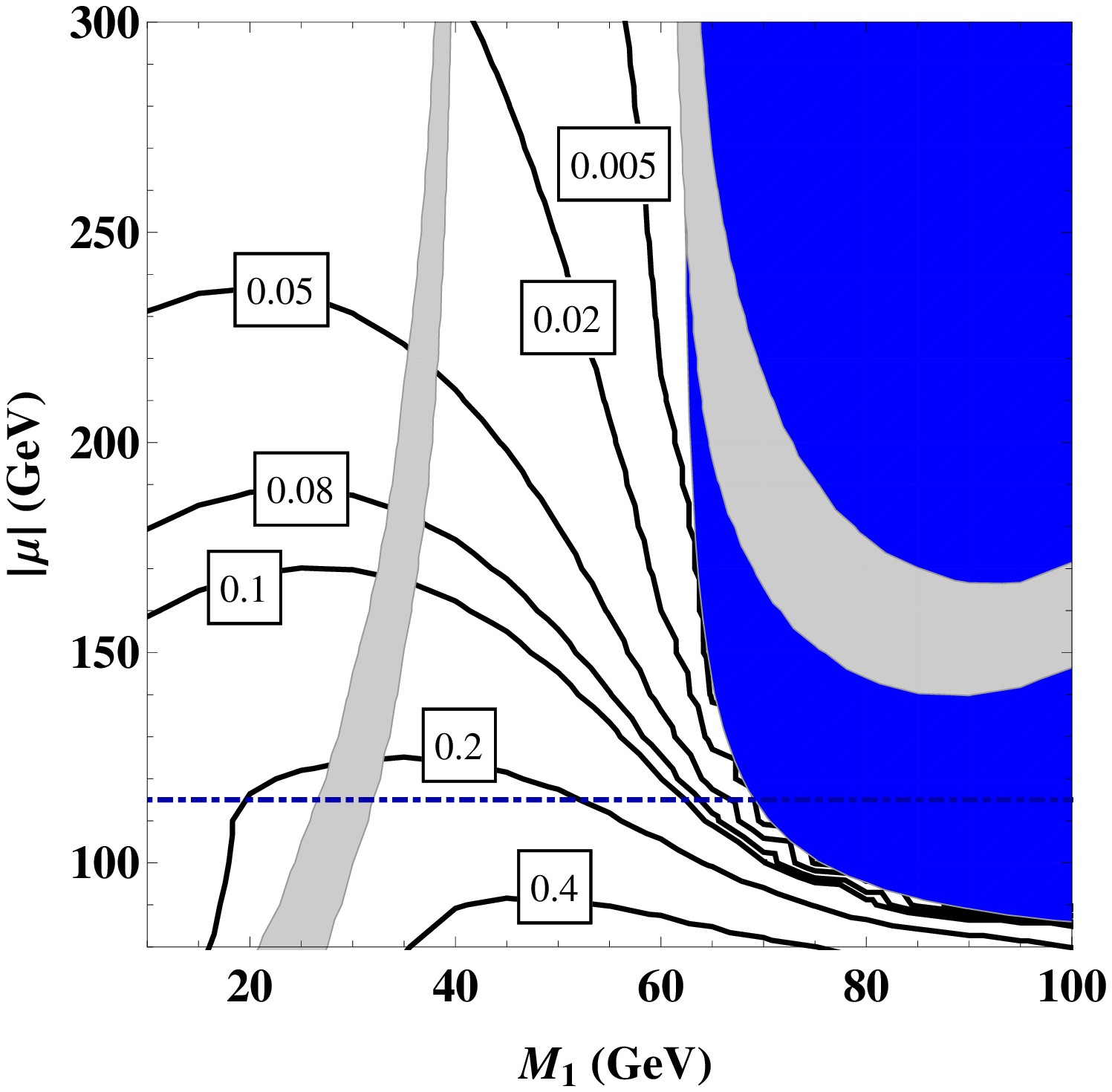}
\hspace{0.3cm}
\includegraphics[width=7.3cm, height=6cm]{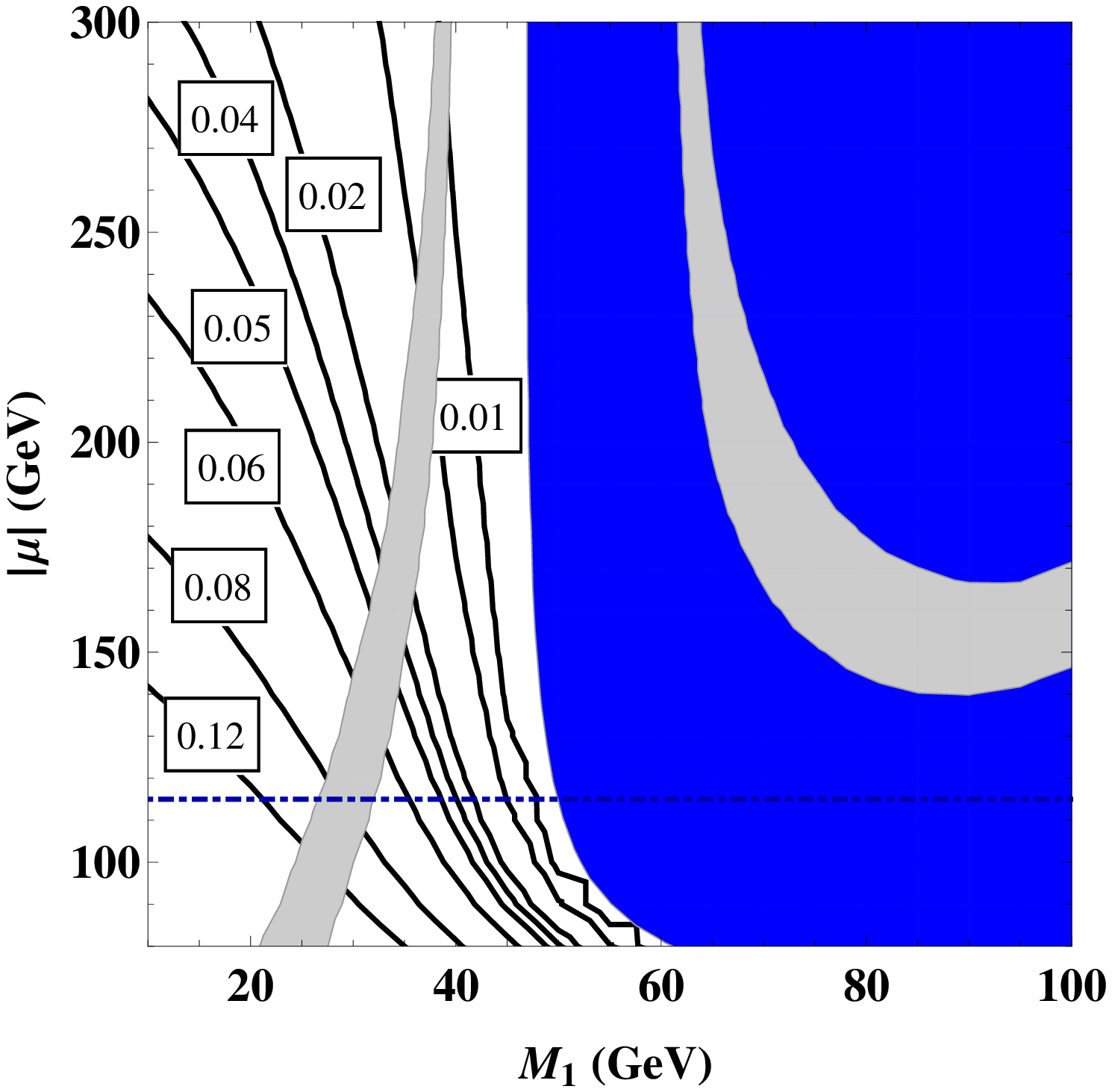}
\caption{The contours of constant branching ratio of ($H \rightarrow \tilde{\chi}_1^0 \tilde{\chi}_1^0$)
(left) and ($h \rightarrow \tilde{\chi}_1^0 \tilde{\chi}_1^0$) (right) 
for a fixed value of $\tan \beta$ = 6 and $M_2$ = 200 GeV with $\mu <$ 0. The region shaded in Blue 
is the region
kinematically not allowed, whereas the Grey shaded region is the one with relic density
within the experimental limits.}
\label{fig:mssm_mu}
\end{figure}

The dependence of our result on the other input parameters is as follows.
If the gaugino mass parameter $M_2$ is lowered, the mass bound of the chargino
pushes up the Blue-DotDashed line in Fig.~\ref{fig:higgs_mssm} to large values of $\mu$
and vice-versa.
The coupling of the Higgs ($h/H$) with neutralino decreases with the increase
of $\tan\beta$ resulting in smaller invisible branching ratio. Large values
of $\tan\beta$ as discussed before are  disfavored in the non-decoupling scenario,
by the LHC experiments. Since there is an enhancement in the branching ratio,
if a smaller value of $\tan \beta$ is chosen, for the sake of completeness
we also quote the results for
$\tan \beta$ = 4. It can be seen from the left hand side of Fig.~\ref{fig:higgs_mssm} 
that for $\tan \beta$ = 6, the largest 
possible branching ratio of $H$ to a pair of neutralinos is 8\%, if the LEP constraint 
on the chargino mass is considered. On the other hand
if the value of $\tan \beta$ is reduced to 4, the maximum possible BR 
satisfying the LEP constraint is 12\%. 
In case of $h$ the change is significant, as for $\tan \beta$ = 6, 
the maximum possible BR was around 28\% as can be seen from the 
right hand side of Fig.~\ref{fig:higgs_mssm}. For $\tan \beta$ = 4, this increases to about 40\%.

We next show in Fig.~\ref{fig:mssm_mu}, the 
contours of the invisible branching ratio of CP even Higgs
$H$ and $h$ to the lightest neutralinos for $\mu$ less than 0. The values of the 
other parameters are same as before with $M_2$ = 200 GeV and $\tan \beta$ = 6.
The chargino mass bound on $\mu$ decreases due to the increase in chargino mass.
The branching ratio of $Z$ to a pair of neutralinos decreases, with negative
$\mu$, therefore the considered $\mu - M_1$ parameter space is not
constrained by Eq.~(\ref{z_width}) in this case.
The neutralino mass also increases with negative $\mu$, resulting in larger
regions of parameter space being kinematically not allowed.
The invisible branching ratio of $H$ increases for negative
$\mu$. This is mainly because for $\mu >$ 0, there is a cancellation
between the terms $Z_{13}\cos \alpha$ and $Z_{14}\sin \alpha$ of (\ref{Higgs_decay1}),
whereas these two terms add for $\mu <$0 leading to enhanced neutralino Higgs coupling.
This behaviour is just the opposite for $h$ as can be seen from the right plot
of Fig.~\ref{fig:mssm_mu}. Here for $\mu <$0, there is a cancellation between the 
$Z_{13}\sin \alpha$ and $ Z_{14}\cos \alpha$ terms of (\ref{higgs_decay1})
leading to reduced coupling of Higgs to the neutralinos.
Overall it can be seen from the above that in the non-decoupling regime
it will not be possible to constrain the neutralino sector from the recent Higgs results 
unlike the decoupling regime where it was possible to do so~\cite{Dreiner:2012ex}.
More data from the LHC is needed so as to constrain the neutralino parameter space
from the Higgs result in this case.

\section{Decay of the Higgs in the NMSSM}\label{sec:nmssm}
We now extend the considerations of the previous section to the NMSSM which has a richer Higgs and 
neutralino sector. To recapitulate, the NMSSM has an extra gauge singlet superfield $S$
in addition to the two Higgs doublets $H_1$ and $H_2$, of the MSSM.
The Higgs(ino) mass term $\mu H_1 H_2$ in the superpotential
of the MSSM is replaced by the trilinear coupling $\lambda S H_1 H_2$,
where $\lambda$ is a dimensionless coupling~\cite{Fayet:1974pd, Ellis:1988er, Drees:1988fc, Pandita:1993hx,
Pandita:1993tg, Ellwanger:1993hn, Elliott:1993uc}. There is also an additional
trilinear self coupling of the singlet $S^3$. The superpotential involving 
only the Higgs field then takes the form
\begin{equation}\label{nmssm_1}
 W_{NMSSM} = \lambda S H_1 H_2 - \frac{\kappa}{3} S^3
\end{equation}
Once the scalar potential of the singlet superfield acquires
a vacuum expectation value $s$, the first term of the superpotential
(\ref{nmssm_1}) then generates an effective $\mu$ term, where 
$\mu_{eff} = \lambda s$. This $\mu_{eff}$ term is naturally
of the order of the electroweak scale, thereby providing a solution to
the $\mu$ problem of MSSM. The Higgs sector of the NMSSM at tree level is described by
six parameters $\mu_{eff},~\lambda,~\kappa,~\tan \beta,~A_\lambda$ and
$A_\kappa$, compared to the Higgs sector of the MSSM, which is only
defined by two independent parameters ($\tan \beta,~M_A$). The physical
Higgs spectrum consists of 3 CP-even states, 2 CP-odd states along
with a pair of charged Higgs boson. The neutralino sector in case of the NMSSM,
due to the addition of the singlet becomes a $5 \times 5$ matrix, which can be written 
in the bino, wino, Higgsino and singlino basis~\cite{Pandita:1994ms, Pandita:1994vw, Choi:2004zx}.
It is described by six independent
parameters $\mu_{eff},~M_1,~M_2,~\tan \beta,~\lambda$ and $\kappa$.
\begin{eqnarray}
\label{nmssmneut}
M_{\mathrm{NMSSM}} = \begin{pmatrix}
M_1 & 0   & - M_Z \sw \cos\beta & \phantom{-}M_Z\sw \sin\beta & 0 \\
0   & M_2 & \phantom{-} M_Z \cw \cos\beta  & -M_Z \cw\sin\beta & 0 \\
- M_Z \sw \cos\beta &\phantom{-} M_Z \cw \cos\beta  & 0 & -\mu_{eff}
& -\lambda v_2\\
\phantom{-}M_Z\sw \sin\beta& -M_Z \cw\sin\beta & -\mu_{eff} & 0 & -\lambda v_1\\
0 & 0 &  -\lambda v_2 & -\lambda v_1 & 2 \kappa x
\end{pmatrix}.
\end{eqnarray}
\noindent
For a massless neutralino the
determinant of the mass matrix  (\ref{nmssmneut})
should be zero, which leads to~\cite{Gogoladze:2002xp}
\begin{eqnarray}
2 \kappa x \mu_{eff} (\Delta_{0} \sin 2 \beta - \mu_{eff} M_1 M_2 )
+ \lambda^2 v^2\left[\Delta_{0}
-\mu_{eff} M_1 M_2 \sin 2 \beta \right] = 0,
\label{det_nmssm}
\end{eqnarray}
\noindent
where $\Delta_{0} = M_{Z}^{2} (M_1 \cos^2 \theta_W + M_2 \sin^2 \theta_W)$.
(\ref{det_nmssm}) in turn leads to the following condition 
\begin{equation}\label{nmssm_mn}
\kappa = \frac{\lambda}{2}\left(\frac{\lambda v}{\mu_{eff}}\right)^2
\frac{\Delta_{0}
-\mu_{eff} M_1M_2 \sin 2\beta}{\mu_{eff} M_1 M_2 -\Delta_{0}\sin 2\beta},
\end{equation}
\noindent
for a massless neutralino in the NMSSM.

Analogous to the MSSM, even in case of the NMSSM it was earlier investigated,
whether the recent global fits from the Higgs data can constrain
the parameter space of the neutralino sector~\cite{Ananthanarayan:2013fga},
with the lightest Higgs ($h_1$) of the NMSSM, being identified as the 126 GeV state observed at LHC. 
There can also be another possibility where the second
lightest CP even Higgs ($h_2$) will lead to a SM like Higgs boson 
in the mass range 124 GeV $\leq M_{h_2} \leq$ 127 GeV. The mass 
of the lightest CP even Higgs $h_1$
and sometimes the lightest pseudoscalar Higgs $a_1$ will be less than $M_{h_2}$, and
in some regions of the parameter space, the decay of $h_2$ to a pair
of $h_1$ or $a_1$ will be kinematically allowed. 
A lot of work has been done in the context of two light Higgs
boson within NMSSM. Various scenarios have been proposed
in this context and are examined or constrained in the light of 
the recent LHC results. 
\begin{enumerate}
 \item[$(a)$]One of them was proposed to explain
the enhancement of the Higgs signal in some of the channels
relative to the SM. The authors of~\cite{Belanger:2012sd,Gunion:2012he,Gunion:2012gc}
have identified a set of parameter space, in the context of NMSSM
where the two lightest CP even Higgs boson are found to be closely
degenerate and lie in the mass window 123-128 GeV. We do not consider this 
possibility here.
\item[$(b)$] Another scenario that has been widely considered in the 
context of NMSSM, is where the heavier Higgs boson $h_2$
is considered as the SM like Higgs in the mass range of
[124, 127] GeV and the lighter Higgs 
boson $h_1$ is around 98 GeV in order to account for
the LEP excess~\cite{Cerdeno:2013cz}. We will refer to this as the 
98 + 126 GeV Higgs scenario further in the text.
\end{enumerate}
Since $h_1$ is in the mass range (96 -100) GeV,
in order to respect the LEP limit of 
$C^{2b}_{eff} = [g^2_{ZZh}/g^2_{ZZh_{SM}}] {\rm BR}(h\rightarrow b\bar{b})$,
from the process $e^+e^- \rightarrow hZ \rightarrow b\bar{b}Z$,
the mass of the lightest pseudoscalar $a_1$ is assumed to be less
than $2 M_b$. There has been additional constraints on the mass of
$a_1$ from various other experiments. In a recent result from CMS,
the experiment has excluded pseudoscalar mass in the range 1 GeV $< M_{a_1} < 2 M_\tau$,
for a scalar Higgs in the mass range 86 - 150 GeV. Therefore
in order to study the 98 + 126 GeV scenario, the light pseudoscalar should 
be either in the range $2 M_\tau < M_{a_1} < 2 M_b$ or heavier than $M_{h_2}$.
Another way of evading the CMS
bound in this two light Higgs scenario, is to consider the mass of the 
lightest scalar ($M_{h_1}$) to be less than 86 GeV~\cite{Chatrchyan:2012cg}. 
The LEP searches of a Higgs boson 
decaying into four $\tau$ leptons via intermediate pseudoscalar~\cite{Schael:2010aw},
has placed constraint on the combined production times branching ratio on the 4 $\tau$'s 
decay channel ($\sigma(e^+e^- \rightarrow Zh)/\sigma_{SM}(e^+e^- \rightarrow Zh))
\times {\rm BR} (h \rightarrow a_1 a_1) \times {\rm BR} (a_1 \rightarrow \tau^+ \tau^-)^2 < 1$). 
All these searches
have mainly considered the decay of Higgs to pseudoscalar as the only
non-standard decay mode apart from the usual SM decay channels. 
Since in our analyses, there are other non-standard decay modes of the Higgs
boson, like the Higgs decaying to a pair
of lightest neutralinos, the constraint on the mass of the pseudoscalar will be
lightened by the presence of these additional decay channels.
We have considered the mass of the pseudoscalar such that the LEP limit 
from the process $e^+e^- \rightarrow hZ,~ h\rightarrow b\bar b$
is satisfied along with the process $e^+e^- \rightarrow hZ,~ h\rightarrow a_1a_1,
~a_1\rightarrow b\bar b$. In addition we have also seen that the LEP constraint
on the 4 $\tau$'s final state is also satisfied.

\begin{figure}[htb]
\begin{minipage}[b]{0.45\linewidth}
\centering
\vspace*{0.7cm}
\includegraphics[width=7cm, height=5cm]{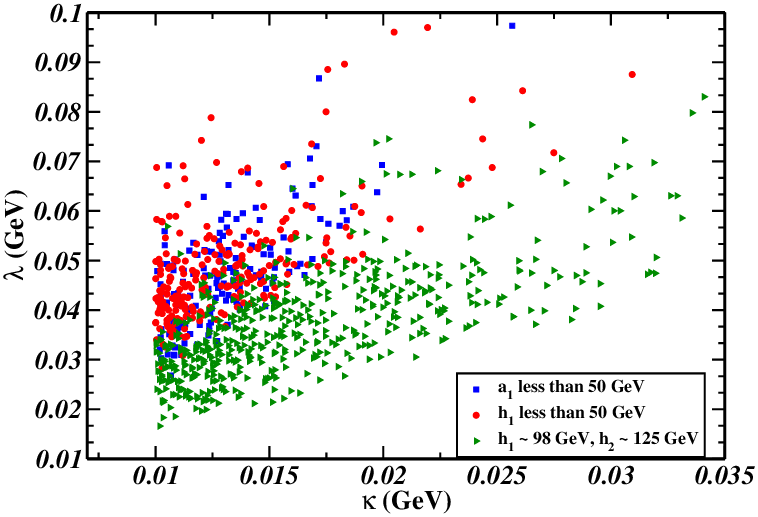}
\caption{Point satisfying all the experimental constraints in the 
$\lambda-\kappa$ plane.}
\label{fig:lam_kap}
\end{minipage}
\hspace{0.4cm}
\begin{minipage}[b]{0.45\linewidth}
\centering
\includegraphics[width=7cm, height=5cm]{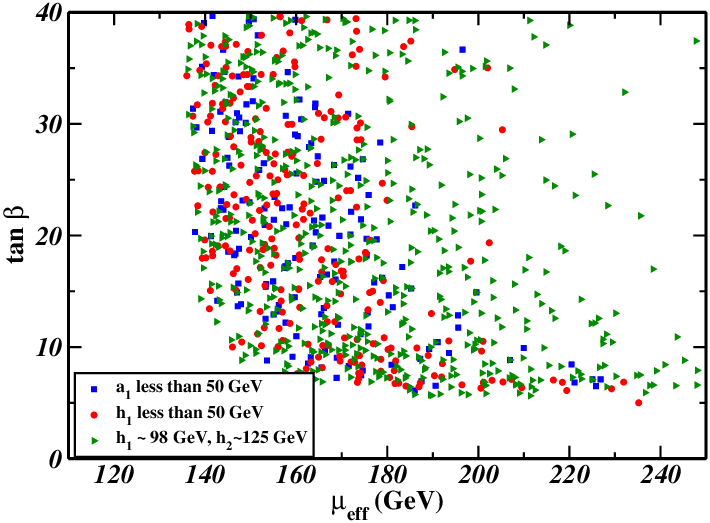}
\caption{Point satisfying all the experimental constraints in the 
$\tan \beta-\mu_{eff}$ plane.}
\label{fig:mu_tanbeta}
\end{minipage}
\begin{minipage}[b]{0.45\linewidth}
\centering
\vspace*{0.7cm}
\includegraphics[width=7cm, height=5cm]{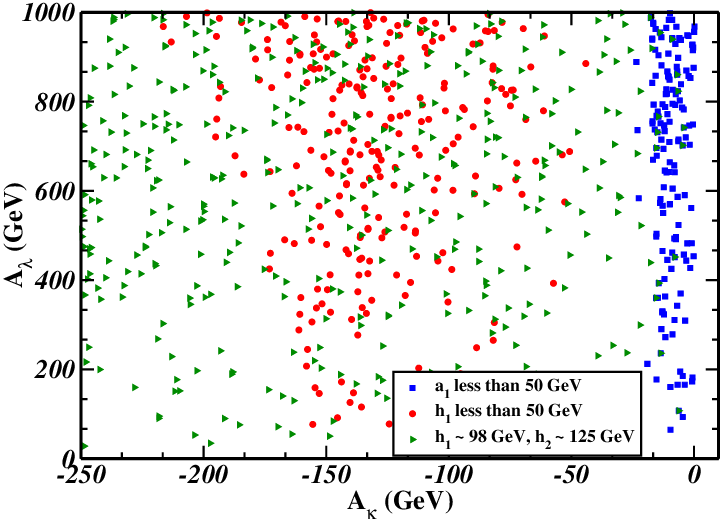}
\caption{Point satisfying all the experimental constraints in the 
$A_\lambda-A_\kappa$ plane.}
\label{fig:akap_alam}
\end{minipage}
\end{figure}

The Higgs sector of the NMSSM being described by six independent parameters,
$\mu_{eff},~\lambda,~\kappa,~\tan \beta,~A_\lambda$ and $A_\kappa$,
a scan is performed over a million random points in the range of parameters
listed in  Table~\ref{tab:nmssm_param}. We have used 
NMSSMTools-4.1.0~\cite{nmssm_web, Ellwanger:2004xm} for our analysis.
The scan includes all the recent experimental constraints from the Higgs,
flavour and precision electroweak measurements implemented within NMSSMTools.
We have additionally demanded that the second lightest CP even Higgs of the NMSSM ($h_2$)
should lead to a SM like Higgs in the mass range [124, 127] GeV. We have also restricted
to values of $\kappa$ and $\lambda$ less than 0.7. This is due to the theoretical
constraint that there should be no charge and color breaking global minima of the scalar
potential and that a Landau pole does not develop below the GUT scale. Since the Higgs mass
spectrum is independent of the gaugino mass parameters, we have considered 
universal boundary conditions at the GUT scale, with the  
$SU(3)_C$ gaugino mass parameter $M_3$ = 1400 GeV, from the gluino searches at the LHC.
The remaining two soft SUSY breaking gaugino parameters have values $M_1$ = 197 GeV
and $M_2$ = 395 GeV. 
\begin{table}
\begin{center}
\begin{tabular}{||c|c|c||}\hline
  parameter & lower range &upper range \\ \hline \hline
$\mu_{eff}$ &100 &400 \\
$\tan \beta$ &5 &40 \\
$\lambda$ &0.01	 &0.7  \\
$\kappa$ &0.01   &0.6  \\
$A_{\lambda}$ &-500 &1000 \\
$A_{\kappa}$ &-1000 &100 \\ \hline 
\end{tabular}
\caption{Ranges of the input parameters of the NMSSM of our scan}
\label{tab:nmssm_param}
\end{center}
\end{table}
In this work, we have divided the points which survive all the constraints defined above
into two distinct scenarios.
\begin{itemize}
\item Scenario 1: The heavier Higgs boson $h_2$ is in the mass range [124, 127] GeV,
whereas the lightest pseudoscalar $a_1$ has a mass less than half the mass of $h_2$.
The lightest CP even Higgs $h_1$ is lighter than $h_2$.
\item Scenario 2 : As before $h_2$ in the mass range [124, 127] GeV,
with the lightest CP even scalar $h_1$ less than half the mass of $h_2$.
The lightest pseudoscalar $a_1$ can be lighter than $h_1$ satisfying the experimental 
constraints or heavier than $h_2$.
\end{itemize}
The 98 + 126 GeV Higgs Boson case, can be obtained in 
the first scenario, but
we separately give the parameter points which satisfy this.
We show in Figs.~\ref{fig:lam_kap},~\ref{fig:mu_tanbeta},~\ref{fig:akap_alam}
the different parameters, that lead to the scenarios of interest considered
here. The points in the plots as discussed before satisfies all the experimental
constraints. It can be seen from Fig.~\ref{fig:lam_kap} that smaller values of 
$\lambda$ and $\kappa$ are preferred for the scenarios we are considering.
A lot of work has been done in the context of large doublet singlet mixing
in the Higgs sector, i.e. concentrating on regions of the parameter space
with large values of $\lambda$, leading naturally to a SM like Higgs $h_2$
in the 126 GeV range~\cite{Ellwanger:2011aa, Cao:2012fz, Ellwanger:2012ke, Beskidt:2013gia}.
But we are mainly interested in the case where the Higgs decay channels
to non-standard particles are open, such as $h_2 \rightarrow a_1a_1, Za_1, h_1h_1$,
along with the neutralinos. Most of the points which satisfy the above constraints
are concentrated in the low $\kappa-\lambda$ plane, therefore we show them here.
Moreover Scenario 1 is distinct from the others
in the $A_\lambda-A_\kappa$ plane, since $M_{a_1}$ is sensitive to $A_\kappa$.
With the universal gaugino masses at the GUT scale, and from (\ref{nmssm_mn})
we find that it is not possible to get a massless neutralino in the NMSSM, 
with $M_{h_2} \approx$ 126 GeV. This result holds in the entire parameter space considered
in our analyses. In the range of the parameter space considered by us,
for NMSSM with universal boundary conditions of gaugino masses at the GUT scale, the 
decay $h_2 \rightarrow \tilde{\chi}_1^0 \tilde{\chi}_1^0$ is kinematically possible in
some regions. Since analysis with the GUT relation between $M_1$ and $M_2$ will result 
in confining to a particular case, we do not consider that possibility here but instead concentrate on the 
general case with $M_1$ and $M_2$ as independent parameters. We consider a benchmark
point for the different scenarios listed above and present the results here. 
The spectrum of the sparticles and the gluinos are considered similar to the case
of MSSM. We list in the Table~\ref{tab:bs_param} the parameters for the two
different benchmark scenarios considered here. The corresponding Higgs spectrum is listed 
in Table~\ref{tab:higgs_spec}.
\begin{table}
\begin{center}
\begin{tabular}{||c|c|c|c|c|c|c|c||}\hline
  &$\lambda$ & $\kappa$ &$A_\lambda$ &$A_\kappa$ &$\tan \beta$ &$\mu_{eff}$ &$A_{t,b,\tau}$\\ \hline \hline
Scenario 1 &0.055 &0.013 &875.76 &-0.174 &19.97 &169.47 &-2500 \\ 
  &&&&&&& \\
Scenario 2 &0.037 &0.013 &978.21 &-168.44 &18.69 &149.77  &-2500 \\ \hline
\end{tabular}
\caption{Input parameters for the Benchmark Points in case of NMSSM}
\label{tab:bs_param}
\end{center}
\end{table}
\begin{table}
\begin{center}
\begin{tabular}{||c|c|c|c|c|c|c||}\hline
  &$M_{h_1}$ & $M_{h_2}$ &$M_{h_3}$ &$M_{a_1}$ &$M_{a_2}$ &$M_{h^{\pm}}$ \\ \hline \hline
Scenario 1 &76.28 &126.47 &1716.62 &5.23 &1716.6 &1718.2 \\ 
&&&&&& \\
Scenario 2 &47.69 &124.58 &1657.66 &164.07 &1657.63 &1659.38 \\ \hline
\end{tabular}
\caption{The Higgs mass spectrum for the different Benchmark Scenarios in case of the NMSSM}
\label{tab:higgs_spec}
\end{center}
\end{table}

\begin{figure}[htb]
\begin{minipage}[b]{0.45\linewidth}
\centering
\vspace*{0.7cm}
\includegraphics[width=7cm, height=5cm]{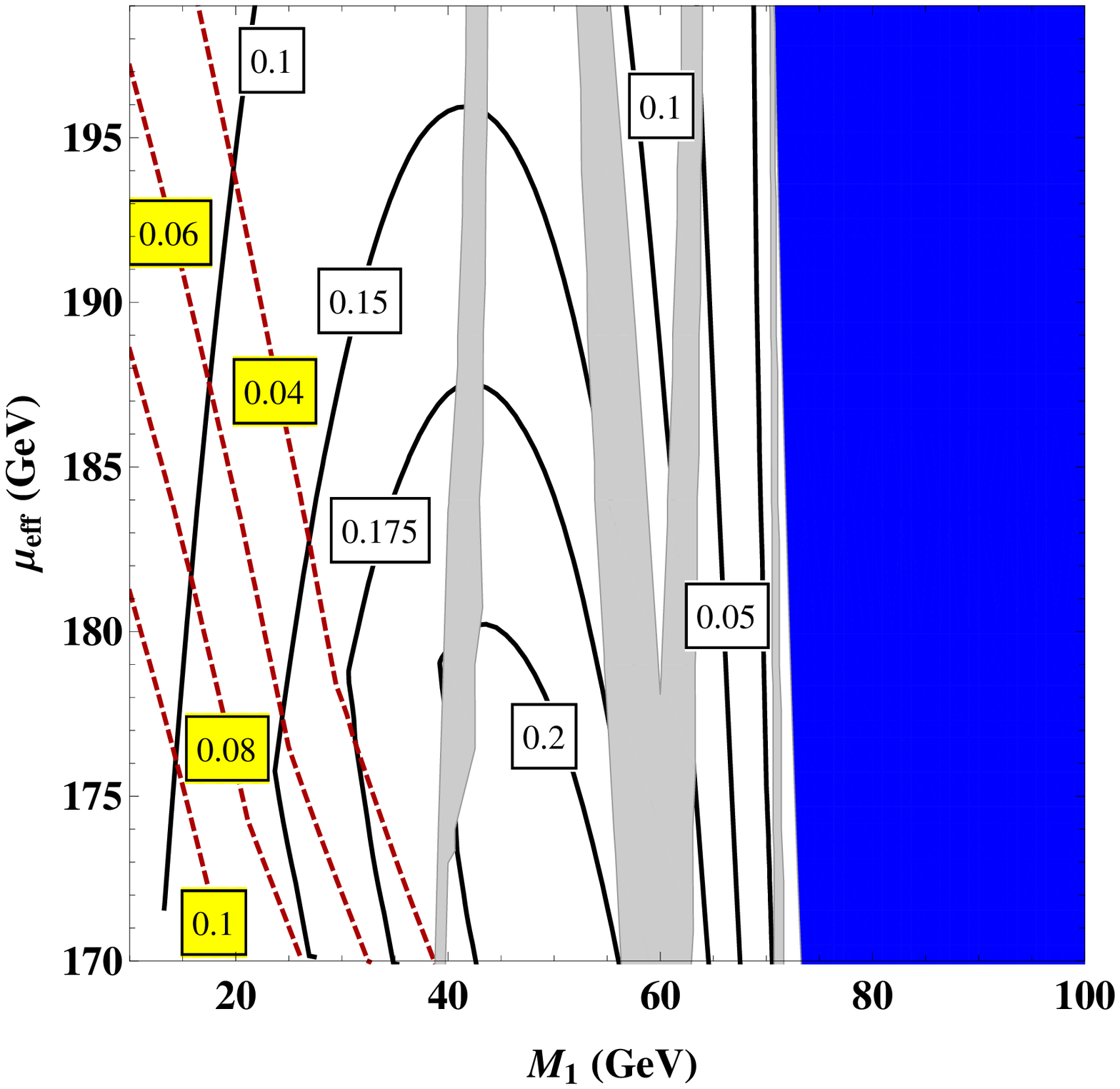}
\caption{Contours of constant branching ratio of 
($h_2\rightarrow \tilde{\chi}_1^0 \tilde{\chi}_1^0$) [Black-Solid]
and ($h_2\rightarrow \tilde{\chi}_1^0 \tilde{\chi}_2^0$) [Red-Dashed]
in NMSSM in the $\mu_{eff}-M_1$ plane for a fixed value of $M_2$ = 200 GeV
and the other parameters fixed to values in Table~\ref{tab:bs_param} for Scenario 1.}
\label{fig:sce1_1}
\end{minipage}
\hspace{0.4cm}
\begin{minipage}[b]{0.45\linewidth}
\centering
\includegraphics[width=7cm, height=5cm]{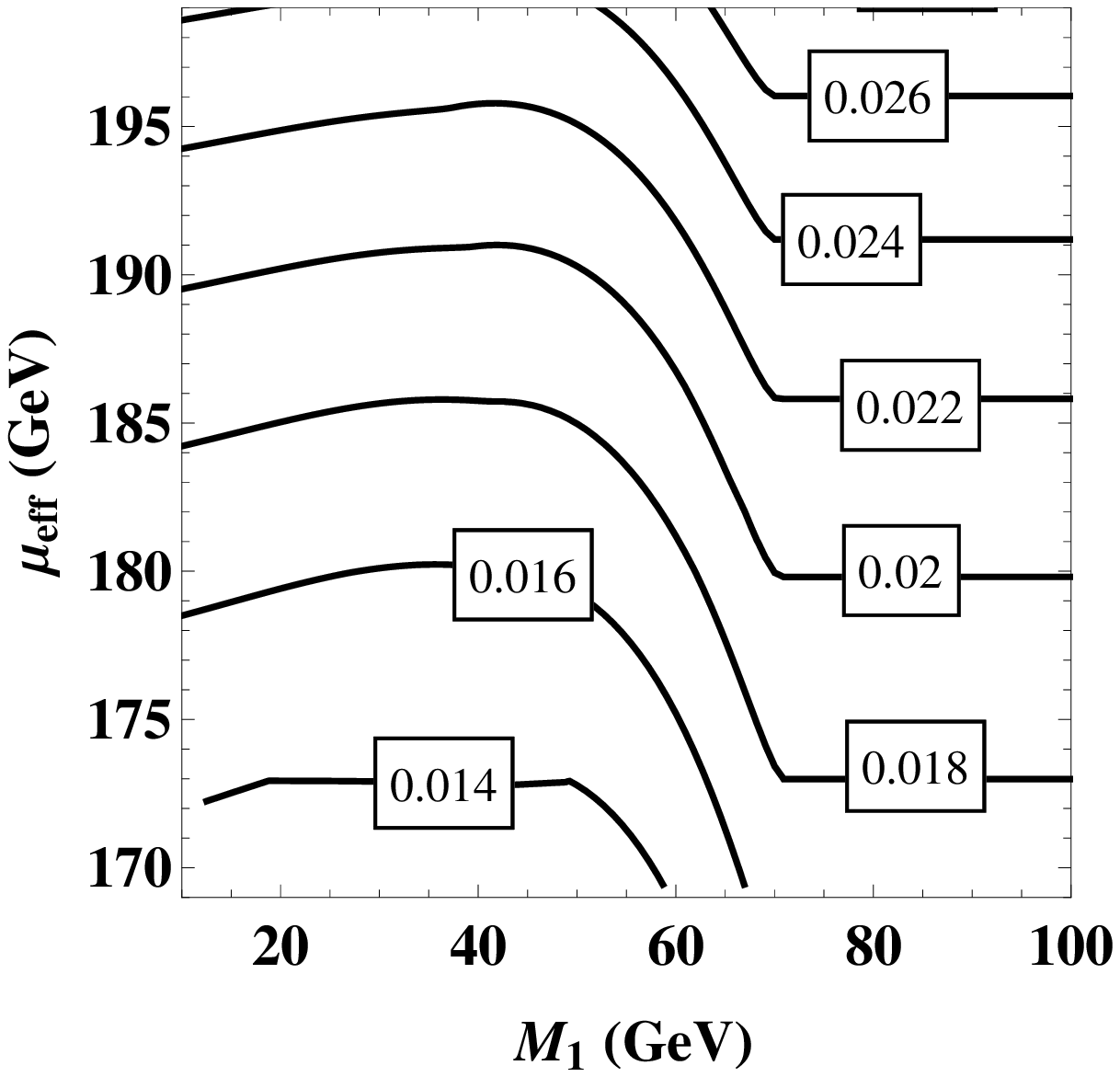}
\caption{Contours of constant branching ratio of ($h_2\rightarrow a_1a_1$)
in NMSSM in the $\mu_{eff}-M_1$ plane for a fixed value of $M_2$ = 200 GeV
and the other parameters fixed to values in Table~\ref{tab:bs_param} for Scenario 1.}
\label{fig:sce1_2}
\end{minipage}
\begin{minipage}[b]{0.45\linewidth}
\centering
\vspace*{0.7cm}
\includegraphics[width=7cm, height=5cm]{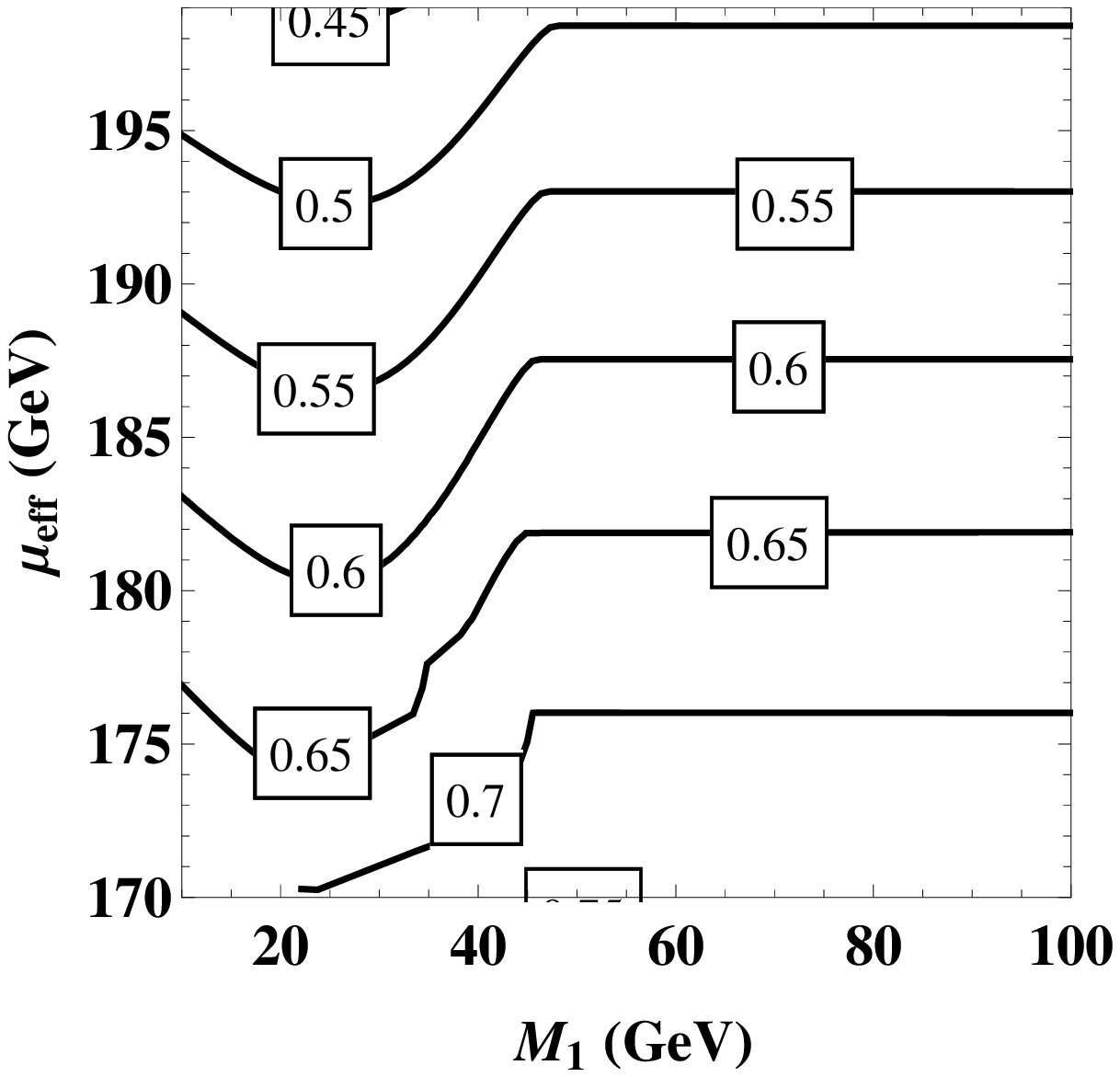}
\caption{Contours of constant branching ratio of ($h_1\rightarrow a_1a_1$)
in NMSSM in the $\mu_{eff}-M_1$ plane for a fixed value of $M_2$ = 200 GeV
and the other parameters fixed to values in Table~\ref{tab:bs_param} for Scenario 1.}
\label{fig:sce1_3}
\end{minipage}
\end{figure}   
We show in Fig.~\ref{fig:sce1_1}, the contours of constant branching ratios
of $h_2 \rightarrow \tilde{\chi}_1^0 \tilde{\chi}_1^0$ (Black)
and $h_2 \rightarrow \tilde{\chi}_1^0 \tilde{\chi}_2^0$ (Red-Dashed)
in the $\mu_{eff}-M_1$ plane with the values of the other parameters
fixed as given in Table~\ref{tab:bs_param} for Scenario 1. Since $M_{h_2}$ 
is sensitive to $\mu_{eff}$, we have varied $\mu_{eff}$ in the range such
that $M_{h_2} \approx$ 124 - 127 GeV.
The blue shaded region is the area where the $h_2$ decay to the lightest neutralinos
is kinematically not accessible. The lightest neutralino has a dominant gaugino component, 
in the entire $\mu_{eff}-M_1$ plane. The singlino component is absent for low 
$M_1$ values, and open up at higher values of $M_1$. 
The grey shaded area shows the region where the 
lightest neutralino satisfies the relic density constraint. 
The relic density is satisfied in the region, where the
neutralino is a gaugino-higgsino mixture, but has a dominant gaugino component.
We see that for the region allowed by the relic density, the invisible branching ratio can vary 
between the range of 15\%-20\%. As invisible BRs less that 38\% is still allowed by the global
fits, the NMSSM parameter space can not be constrained by the present LHC Higgs 
data. In the future with the upgraded LHC results, it will be possible 
to constrain the parameter space
from the Higgs data. The nature of the contours in Fig.~\ref{fig:sce1_1} can be readily
understood from the fact that, since $M_{\tilde{\chi}_2^0}$ also depends on
$\mu_{eff}$ and $M_1$, at low values of $M_1$, the decay channel 
$h_2 \rightarrow \tilde{\chi}_1^0 \tilde{\chi}_2^0$
is kinematically accessible. This is shown by Red-Dashed lines which decreases
with increasing $M_1$ due to the increase in mass of $\tilde{\chi}_2^0$,
and the opening of the channel $h_2 \rightarrow \tilde{\chi}_1^0 \tilde{\chi}_1^0$.
The second lightest neutralino is mostly a singlino.
Since in this scenario the lightest pseudoscalar is very light, $M_{a_1} \approx$ 6 GeV,
we show in Fig.~\ref{fig:sce1_2}, the branching ratio of $h_2$ to a pair of $a_1$
in the $\mu_{eff}-M_1$ plane. The decay channel $h_2 \rightarrow Z a_1$ is also open.
but the branching ratio is very small. So we do not consider it here. The nature of the 
contour in Fig.~\ref{fig:sce1_2} can be understood from the fact that at high values of $M_1$
since the neutralino decay channel is not there, the BR is constant. Whereas for lower values
of $M_1$ due to the invisible branching ratio, the contours show a curved nature. Here $a_1$
will mostly decay into a pair of $\tau$'s, which can be easily detected in the collider. So we do not
include them in the calculation of the invisible branching ratio. 
We also show in Fig.~\ref{fig:sce1_3} the decay of the light CP even Higgs $h_1$
to a pair of $a_1$. Here $h_1$ predominantly decays to $a_1$ with around 50-60\% branching ratio.
The dip in the contours at lower values of $M_1$ is due to the presence of light neutralinos,
leading to the decay channel $h_1 \rightarrow \tilde{\chi}_1^0 \tilde{\chi}_1^0$.
The lighter Higgs $h_1$ in this case can be observed in the collider through the 
decay mode $h_1 \rightarrow a_1a_1\rightarrow 4\tau$'s.
\begin{figure}[htb]
\begin{minipage}[b]{0.45\linewidth}
\centering
\vspace*{0.7cm}
\includegraphics[width=7cm, height=5cm]{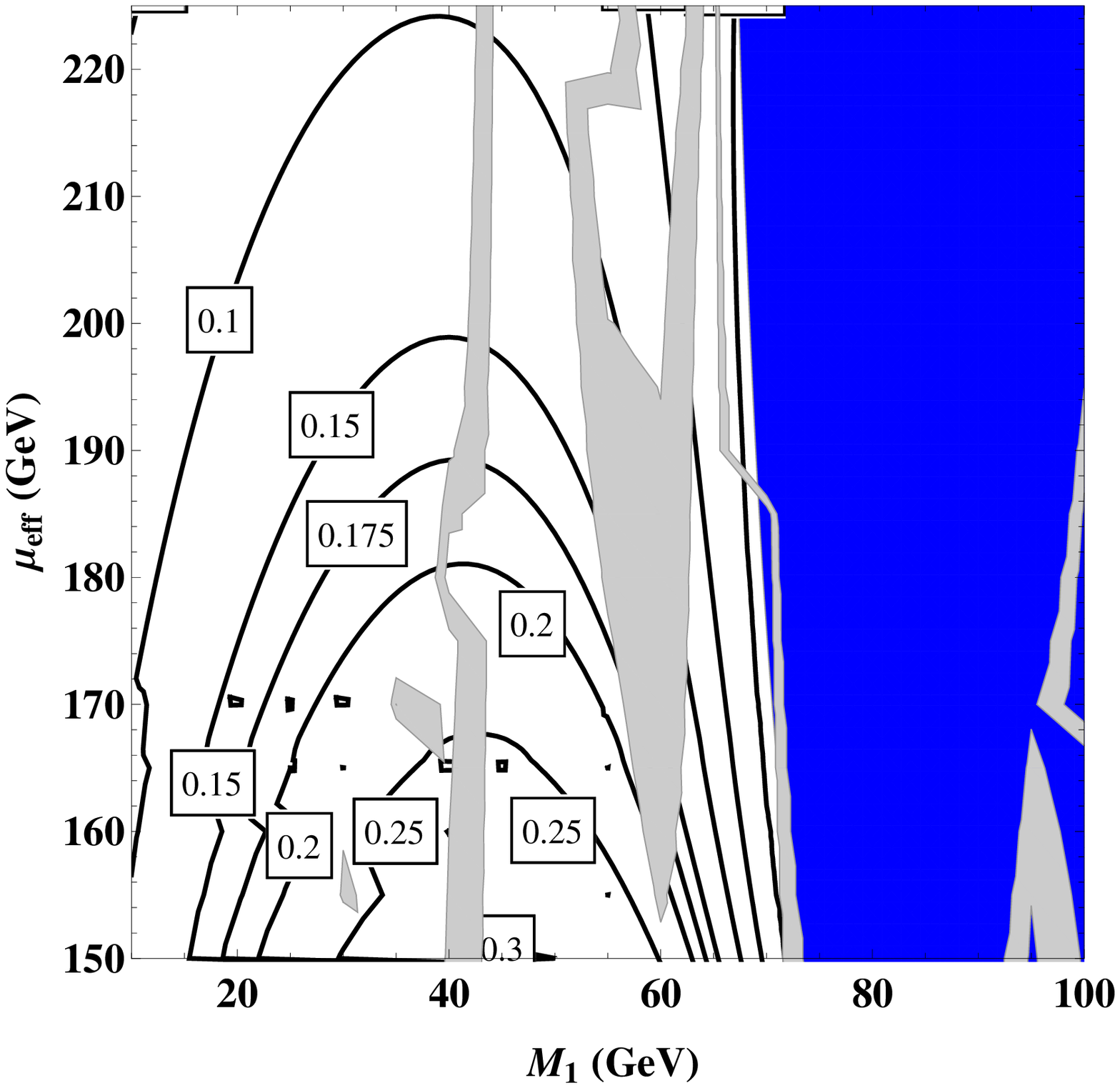}
\caption{Contours of constant branching ratio of 
($h_2\rightarrow \tilde{\chi}_1^0 \tilde{\chi}_1^0$) 
in NMSSM in the $\mu_{eff}-M_1$ plane for a fixed value of $M_2$ = 200 GeV
and the other parameters fixed to values in Table~\ref{tab:bs_param} for Scenario 2.}
\label{fig:sce2_1}
\end{minipage}
\hspace{0.4cm}
\begin{minipage}[b]{0.45\linewidth}
\centering
\includegraphics[width=7cm, height=5cm]{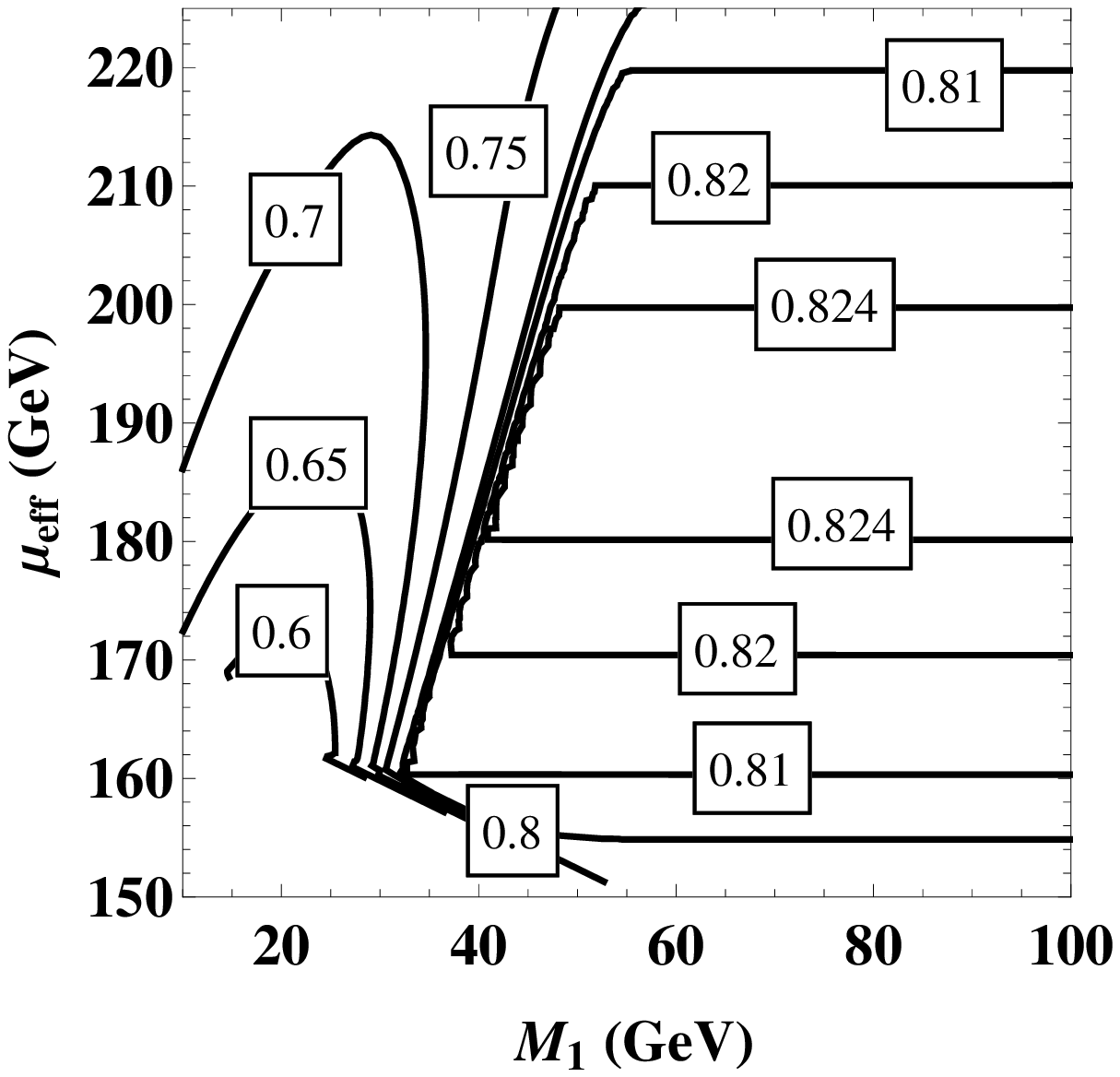}
\caption{Contours of constant branching ratio of 
($h_1\rightarrow b \bar{b}$) in NMSSM in the $\mu_{eff}-M_1$ plane for
a fixed value of $M_2$ = 200 GeV
and the other parameters fixed to values in Table~\ref{tab:bs_param} for Scenario 2.}
\label{fig:sce2_2}
\end{minipage}
\end{figure}
We next show our results for Scenario 2, where $M_{h_1} \leq M_{h_2}/2$.
The nature of the Fig.~\ref{fig:sce2_1} is similar to Fig.~\ref{fig:sce1_1}
showing the contours of constant branching ratios
of $h_2 \rightarrow \tilde{\chi}_1^0 \tilde{\chi}_1^0$.
The value of $\lambda$ in Scenario 2 is smaller compared to
that of Scenario 1, as can be seen from Table~\ref{tab:bs_param}. Since the 
neutralino mass ($M_{\tilde{\chi}_{1,2}^0}$) increases with decreasing
$\lambda$, the neutralinos in this case are more massive. Therefore
the decay channel $h_2 \rightarrow \tilde{\chi}_1^0 \tilde{\chi}_2^0$
is kinematically not accessible. The allowed parameter space can only be constrained 
by the future LHC results. We would further add that the composition of the
lightest and the second lightest neutralino is similar to
Scenario 1. Finally we show in Fig.~\ref{fig:sce2_2}, the contours
of constant branching ratio of the lightest CP even Higgs to a pair of $b$ 
quarks, which is the dominant decay mode. The BR can be as high as 80\% for
most of the $\mu_{eff}-M_1$ parameter space. When the neutralino is light enough
allowing for the invisible decay mode of the light Higgs ($h_1$), the BR
decrease by about 20\%.  It will be possible to observe this state at the 
LHC, through the $b\bar{b}$ decay mode.
The lightest pseudoscalar Higgs in this scenario is heavier
than $h_1$ and $h_2$, but is considerably lighter to be observed at the LHC.
Nevertheless the dominant decay mode of $a$ will be to a pair of neutralinos,
and the second dominant mode will be to a pair of $b$ quarks. The 
pseudoscalar $a$ will therefore be difficult to be observed in the LHC, due to a 
large invisible BR. This shows that the global fits from the recent LHC data is
unable to constrain the neutralino sector of the NMSSM, when the second lightest 
scalar is identified as the 126 GeV Higgs. 
We would like to add that the analysis carried out in this work can
also be repeated for negative values of $\kappa$. However, we have found that
in this case for most of the parameter space with $h_2$ around 126 GeV,
the limits on the invisible branching ratio cannot constrain the parameter
space as is the case with positive values of $\kappa$.
A more detailed precision study in the 
Higgs sector is allowed so as to constrain the parameter space in this case.

\section{Conclusion}\label{sec:conclusion}

In this work we have considered the possibility of the invisible decays of the second lightest
CP even Higgs boson in the context of both MSSM and NMSSM. The second lightest Higgs
behaves as the SM like Higgs with mass in the allowed mass range $\approx$ [124, 127] GeV.
The neutralino sector in the last few years has been studied in details and also constrained
by the data from different astrophysical, cosmological and collider experiments. The recent
LHC results on the Higgs branching ratio has independently constrained the Higgs invisible branching 
ratio, through global fits. 
The Higgs can decay to a pair of neutralinos giving rise to invisible decay branching ratio provided
$M_{\tilde{\chi}_1^0} < M_H$. This invisible Higgs decay channel is also looked through direct searches
at the LHC. It will therefore be important to consider the implications of the current information on
Higgs from the LHC, on the neutralino sector of the supersymmetric models. 
In the context of MSSM, we discuss in brief
whether it will be possible for the SM like Higgs to decay to a pair of neutralinos,
with both universal and nonuniversal gaugino masses at the GUT scale.
We find that with universal gaugino masses, the invisible decay channel 
$H\rightarrow \tilde{\chi}_1^0 \tilde{\chi}_1^0$ is kinematically not allowed.
Nevertheless in case of nonuniversal gaugino masses, for certain representations
of $SO(10)$ and $E_6$, the 126 GeV Higgs will have a considerable invisible branching ratio.
We have then analyzed the possibility of having a large invisible branching ratio
in the context of NMSSM which has a richer neutralino and Higgs sector
compared to the MSSM. We find that in case of the NMSSM it is possible to have the invisible decay channel
($h_2 \rightarrow \tilde{\chi}_1^0 \tilde{\chi}_1^0$) with universal gaugino masses at the
GUT scale.  The assumption of the GUT relation between $M_1$ and $M_2$ being largely
model dependent, we have considered a more general case and have done our analyses
by treating $M_1$ and $M_2$ as two independent parameters in case of both the MSSM and the NMSSM. 
With this assumption, there is an additional freedom of the neutralino being very light.

We have then studied the decay of the Higgs bosons ($h$ and $H$) to a pair of lightest
neutralinos in the context of the non-decoupling scenario of the MSSM. 
We have considered the possibility that the 
126 GeV scalar ($H$) observed at the LHC along with the 98 GeV scalar ($h$) from the LEP excess 
in the $b\bar{b}$ final state can be concurrently explained in the MSSM framework.
We find that there are regions in the MSSM parameter space where such scenarios exist.
The neutralino sector being dependent on the parameters $\mu,~M_1,~M_2,~\tan\beta$,
we give our results for a fixed value of $\tan \beta$ and $M_2$ in the 
$\mu - M_1$ plane because we are mainly interested in the Higgs decay channel to a pair of neutralinos.
We find the invisible BR in this case is too small to be constrained by the recent
LHC fits from the Higgs data. This is in contrast to the case when the lightest CP even Higgs boson 
($h$) was SM like~\cite{Ananthanarayan:2013fga, Dreiner:2012ex}, the decoupling scenario.
There it was found that with $h \approx$ 126 GeV,  a large portion of the 
$\mu - M_1$ parameter space for a fixed value of $\tan \beta$ and $M_2$ allowed 
a large invisible branching ratio in conflict with the latest LHC fits from the Higgs sector,
thereby constraining the neutralino parameter space.  Therefore if the non-decoupling scenario
exists, the parameter space of the neutralino sector cannot be constrained by the recent LHC data.
Higher precision Higgs physics is required  to constrain the parameter space in the scenario.
This scenario can be alternatively tested by looking for the other Higgs bosons production, 
$M_h \approx$ 98 GeV, $M_A \approx$ 100 - 150 GeV and $M_{H^\pm} \approx$ 150 - 200 GeV at the LHC.
We do not consider the possibility here. The dependence of our results on the other input parameters 
is also discussed in details.

An analogous analysis is then performed in the context of the NMSSM. The number of independent parameters
in the neutralino and the Higgs sector is greater than that of the MSSM.  We have
performed a scan over the parameters contributing to the neutralino sector of the NMSSM and 
have plotted the points which pass the various theoretical and experimental constraints discussed in the text,
along with the condition that the second lightest Higgs $h_2$ behaves like the SM Higgs boson. 
The points which survive these constraints are then divided into three scenarios,
depending on the mass of the other Higgs bosons. 
We firstly separately isolate the points where both $M_{h_1,a_1}$ is less than $m_{h_2}$, but only the 
decay of $h_2$ to a pair of CP odd Higgs $a_1$ is kinematically allowed.
Secondly we consider the case when 
the mass of the lightest Higgs boson ($h_1$) is less than half the mass of $h_2$, that is the decay 
$h_2\rightarrow h_1h_1$ is kinematically allowed.
We have considered a single benchmark point for these two scenarios and have exclusively worked out 
the results. The dependence of the results on the other points can be suitably interpreted. 
There are regions in the parameter space which also satisfies the 98 ($h_1$) + 126 ($h_2$) GeV 
Higgs scenario, so as to account for the LEP excess along with the Higgs data from the LHC.
We list the points which fall in this parameter space, but we do not consider benchmark 
point in this case as the results will be similar to the first scenario. 
We see that the invisible branching ratio reaches a maximum of 20\%, for the two scenarios
considered here, still too small to be be constrained by the recent LHC fits from the Higgs 
data.  The result is thus similar to MSSM,
where with the second lightest scalar being 126 GeV, it is currently not possible to constrain the neutralino
parameter space from the Higgs data. The situation will improve with more data from the next LHC run. 

Overall we find that in the context of both MSSM and NMSSM, it is not possible to 
constrain the neutralino sector from the recent Higgs data, if the second lightest scalar
is identified with the one observed at the LHC.
The presence of the other light Higgs will however lead to interesting collider signals at the LHC, 
which will alternatively test these scenarios. 
It will be worthwhile to do a detailed collider study of these light Higgs and 
find the reach of the LHC. The direct detection of some susy particles 
in the 13 and 14 TeV runs of the LHC would significantly cut down the arbitrariness of extensions
of the SM to its susy variants after which a precision Higgs era could be pursued at the LHC.
Hopefully this can be pursued in the near future.

\section{acknowledgements} 
We would like to thank B. Ananthanarayan for a careful reading of the manuscript
and constructive suggestions. MP would also like to thank Dipan Sengupta
for help regarding micrOMEGAs. The work of PNP is supported by the J. C. Bose National Fellowship 
of the Department of Science and Technology, India, and by the Council of Scientific and 
Industrial Research, India.   


\end{document}